\documentclass[aps,prx,twocolumn,superscriptaddress,floatfix,noeprint]{revtex4-2}
\usepackage[utf8]{inputenc}
\usepackage[colorlinks=true,allcolors=blue]{hyperref}

\usepackage{graphicx}
\usepackage{soul}
\usepackage{xcolor}
\graphicspath{{figures/}}

\usepackage{amsmath}
\usepackage{amssymb}
\usepackage{bm}
\usepackage{mathtools}
\usepackage{braket}

\renewcommand{\vec}[1]{{\bm{#1}}}
\DeclarePairedDelimiter{\mean}{\langle}{\rangle}
\DeclarePairedDelimiter{\abs}{\lvert}{\rvert}
\DeclarePairedDelimiter{\norm}{\lVert}{\rVert}
\DeclareMathOperator{\sign}{sign}
\DeclareMathOperator{\tr}{tr}
\DeclareMathOperator{\re}{Re}
\DeclareMathOperator{\im}{Im}
\DeclareMathOperator{\Ci}{Ci}
\DeclareMathOperator{\Si}{Si}

\begin{document}
\setstcolor{purple} 

\title{Fermionic matter-wave quantum optics with cold-atom impurity models}

\author{Bennet Windt}
\email{bennet.windt@mpq.mpg.de}
\address{Max Planck Institute of Quantum Optics, Hans-Kopfermann-Stra{\ss}e 1, 85748 Garching, Germany}
\address{Munich Center for Quantum Science and Technology, Schellingstra{\ss}e 4, 80799 M{\"u}nchen, Germany}

\author{Miguel Bello}
\address{Max Planck Institute of Quantum Optics, Hans-Kopfermann-Stra{\ss}e 1, 85748 Garching, Germany}
\address{Munich Center for Quantum Science and Technology, Schellingstra{\ss}e 4, 80799 M{\"u}nchen, Germany}

\author{Eugene Demler}
\address{Institute for Theoretical Physics, ETH Zurich, 8093 Zurich, Switzerland}

\author{J. Ignacio Cirac}
\address{Max Planck Institute of Quantum Optics, Hans-Kopfermann-Stra{\ss}e 1, 85748 Garching, Germany}
\address{Munich Center for Quantum Science and Technology, Schellingstra{\ss}e 4, 80799 M{\"u}nchen, Germany}

\begin{abstract}
Motivated by recent cold-atom realisations of matter-wave waveguide QED, we study simple fermionic impurity models and discuss fermionic analogues of several paradigmatic phenomena in quantum optics, including formation of non-trivial bound states, (matter-wave) emission dynamics, and collective dissipation. For a single impurity, we highlight interesting ground-state features, focusing in particular on real-space signatures of an emergent length scale associated with an impurity screening cloud. We also present novel non-Markovian many-body effects in the quench dynamics of single- and multiple-impurity systems, including fractional decay around the Fermi level and multi-excitation population trapping due to bound states in the continuum.
\end{abstract}

\maketitle

\section{Introduction}

The quintessential quantum optical system comprises one or multiple few-level quantum systems (e.g.\  atoms) coupled to a dissipative environment (e.g.\ the electromagnetic field). In the context of modern quantum technologies, there has been a particular surge of interest in systems of quantum emitters in \emph{structured} baths, i.e.\ environments with non-trivial spectral properties such as band gaps~\cite{lambropoulos_fundamental_2000}, Van Hove singularities~\cite{gonzalez-tudela_markovian_2017,gonzalez-tudela_non-markovian_2018,gonzalez-tudela_quantum_2017, gonzalez-tudela_anisotropic_2019}, or Dirac cones~\cite{gonzalez-tudela_exotic_2018, perczel_theory_2020, redondo-yuste_quantum_2021}, and even topological features~\cite{bello_unconventional_2019, kim_quantum_2021, bello_topological_2022, vega_qubit-photon_2021, de_bernardis_chiral_2023}. 
Such systems display a host of intriguing physical phenomena. For instance, coupling a single excited emitter to a bath with a gapped band structure can lead to fractional decay due to population trapping in an emitter-bath bound state~\cite{lambropoulos_fundamental_2000, john_quantum_1990, john_spontaneous_1994}, as observed in recent experiments~\cite{liu_quantum_2017, sundaresan_interacting_2019, stewart_dynamics_2020}. Moreover, such bound states can mediate purely coherent long-range interactions between multiple emitters, enabling the simulation of diverse quantum many-body models~\cite{hood_atomatom_2016, douglas_quantum_2015, bello_spin_2022, gonzalez-tudela_subwavelength_2015, munro_optical_2017, zhang_superconducting_2023}.  Structured baths also give rise to a variety of novel dissipative phenomena, including non-exponential decay and chiral emission dynamics~\cite{lodahl_chiral_2017,gonzalez-tudela_markovian_2017,gonzalez-tudela_non-markovian_2018,gonzalez-tudela_quantum_2017, gonzalez-tudela_anisotropic_2019, gonzalez-tudela_exotic_2018, perczel_theory_2020, redondo-yuste_quantum_2021}, as well as collective effects, such as sub- and superradiance~\cite{goban_superradiance_2015, liedl_observation_2022,tiranov_collective_2023,cardenas-lopez_many-body_2023}. 

Crucially, the study of quantum emitters in structured baths has been largely restricted to single-particle physics, with only few attempts to move to the multi-excitation regime~\cite{calajo_atom-field_2016, mahmoodian_dynamics_2020, shi_bound_2016}. Consequently, very little is known about the many-body physics of such models. In particular, the role of particle statistics (i.e.\ the difference between fermionic and bosonic particles) has, to our knowledge, not been thoroughly examined. In the field of condensed matter physics, \emph{fermionic} impurity models are well-established and play a significant role in a wide variety of systems. For instance, individual localised fermionic impurity levels can affect transport properties of mesoscopic systems~\cite{matveev_interaction-induced_1992,geim_fermi-edge_1994,blanter_shot_2000,aleiner_quantum}, and their presence in superconductors can lead to the emergence of localised quasi-particle states~\cite{balatsky_impurity-induced_2006} or Majorana modes~\cite{nadj-perge_proposal_2013}. Fermionic impurity models have also served in the description of the orthogonality catastrophe in x-ray spectroscopy~\cite{ohtaka_theory_1990} as well as the Kondo effect~\cite{Hewson1993}.

By contrast, such models have not received much attention from the quantum optics community, perhaps due to a perceived lack of applicability. However, recent proposals~\cite{de_vega_matter-wave_2008, gonzalez-tudela_non-markovian_2018, navarrete-benlloch_simulating_2011} and subsequent realisations~\cite{stewart_dynamics_2020,krinner_spontaneous_2018} of cold-atom matter-wave analogues to traditional waveguide QED setups provide a quite natural pathway to exploring these models from a quantum optics perspective: by trading  in the bosonic atoms used in these setups for fermionic ones, such experiments could unlock a new scenario of ``fermionic matter-wave quantum optics''. 

In this work, we explore this bridge between the fields of quantum optics and condensed matter physics in the context of a system of fermionic impurities coupled to a structured bath. Specifically, we model the bath as a lattice of non-interacting spinless fermions and the impurities as additional fermionic modes coupled by local hopping to different sites of the lattice. We analyze separately the case of a single impurity and the case of multiple impurities. In either case, we consider both one-dimensional (1D) and two-dimensional (2D) lattices, and we analyze in detail the thermodynamic limit, in which the lattice becomes infinite. 

The single-impurity version of our model, known as the \emph{Resonant Level Model} (RLM)~\cite{Haug2008}, has been extensively studied in 1D, but remains largely unexplored in 2D beyond the effectively 1D description obtained by assuming that the impurity level couples only to $l=0$ bath modes. We explore the ground-state features of this model, incorporating both the effects of a finite bandwidth in the bath and a 2D lattice without full rotational symmetry, and present signatures of a screening cloud in the scaling of impurity-bath correlations as a function of distance $r$ from the impurity. Specifically, in 1D we observe a cross-over from logarithmic scaling to the characteristic $r^{-1}$ scaling of correlations in a bath without an impurity, consistent with previous findings~\cite{ghosh_real-space_2014, schuch_matrix_2019}. However, we show that this scaling is universal over a much wider range of parameter values than previously believed. In 2D, we show that the long-range scaling of the correlations is also the same as that found in a bath without an impurity, of the form $r^{-3/2}$, contrary to the behaviour predicted in the literature~\cite{ghosh_real-space_2014, schuch_matrix_2019}.

We also study the quench dynamics of an initial state in which the impurity modes are occupied and the bath is in its Fermi sea ground state. We use a master equation to describe the dynamics in the Markovian regime~\cite{schuetz_superradiance-like_2012}. To describe non-Markovian effects, we adapt the single-excitation resolvent formalism from quantum optics to the multi-excitation sector of our model. We then discuss the circumstances under which the system relaxes to the ground state, and clarify the role of the single-particle bound states of the system alluded to above in this many-body setting. In particular, for a single impurity, we show that fractional decay occurs not only at the band edges, but also around the Fermi level, and how the emission dynamics differs depending on the value of the Fermi level relative to the impurity on-site potential. In the multiple-impurity case, we show that the dynamics generally follow a multi-exponential decay, which is fractional in certain impurity configurations supporting bound states in the continuum (BICs), and we generalise the theory of BICs in 2D baths presented in Refs.~\cite{gonzalez-tudela_markovian_2017, gonzalez-tudela_quantum_2017}.

Finally, we discuss the feasibility of observing some of these effects in state-of-the-art cold-atom experiments. In particular, we show that the signatures of the ground-state screening cloud can be detected in finite systems, with sizes that are readily achievable in experiments, and we suggest how to prepare the many-body ground state, using an adiabatic or a dynamical protocol. We also discuss the effects of finite temperature on the observation of non-Markovian dynamical phenomena.

The paper is structured as follows: In Sec.~\ref{sec:formalism}, we introduce our model and develop the theoretical tools used throughout. We then discuss the physics of a single impurity (Sec.~\ref{sec:singleimpurity}) and multiple impurities (Sec.~\ref{sec:multipleimpurities}), before addressing experimental factors in Sec.~\ref{sec:experiment}.

\section{Theoretical framework}
\label{sec:formalism}

In this section, we introduce our model and the notation and conventions employed throughout this work. We briefly review the theory of free-fermion models and fermionic Gaussian states and introduce both a perturbative and an exact method for studying the dynamics of our model, based on a Markovian master equation and an extension the resolvent formalism, respectively.

\subsection{Fermionic impurity models}
We consider $N$ spinless fermionic impurities, described by creation and annihilation operators $\{c_n^\dagger, c_n\}_{n=1}^N$, coupled via local tunneling to a $d$-dimensional spinless fermionic bath of $L^d$ sites. 
The bath modes are likewise described by operators $\{b_j^\dagger, b_j\}_{j=1}^{L^d}$, and we denote the site to which the $n$-th impurity is coupled by $j_n$. We restrict our attention to the 1D ($d = 1$) and 2D ($d = 2$) cases. 
The Hamiltonian for this system can be written as $H = H_S + H_B + V$, with ($\hbar=1$)
\begin{subequations}
\label{eq:Hamiltonian}
\begin{align}
    & H_S = \Delta \sum_n c^\dag_n c_n \,, \\
    & H_B = 2^d J \sum_j b^\dag_j b_j - J \sum_{\mean{j, j'}} \left(b^\dag_j b_{j'} + \mathrm{H.c.}\right)\,, \\
    & V = g \sum_n \left(c_n^\dagger b_{j_n} + \mathrm{H.c.}\right) \,.
\end{align}
\end{subequations}
Since the bath is translation-invariant, and it has a single site per unit cell, it can be diagonalised by a Fourier transform. 
Defining modes $b_\vec{k} = \sum_j e^{-i\vec{k}\cdot\vec{r}_j} b_j/\sqrt{L^d}$ with quasi-momenta in the first Brillouin zone (BZ), $\vec{k} \in (2\pi/L)\mathbb{Z}_L^d$, we can express 
\begin{subequations}
\label{eq:Hamiltonian_momentum}
\begin{align}
    & H_B = \sum_\vec{k} \omega_\vec{k} b^\dag_\vec{k} b_\vec{k} \,, \\
    & V = \frac{g}{\sqrt{L^d}} \sum_{n,\,\vec{k}} \left(e^{i\vec{k}\cdot\vec{r}_n} c^\dag_n b_\vec{k} + \mathrm{H.c.}\right) \,,
\end{align}
\end{subequations}
where we use the shorthand notation $\vec{r}_n \equiv \vec{r}_{j_n}$, with $\vec{r}_j$ denoting the position of the $j$-th bath site. 
The dispersion relation in the bath takes the form $\omega_\vec{k} = 2J(1 - \cos k)$ or $\omega_\vec{k} = 2J(2 - \cos k_x - \cos k_y)$ for $d = 1$ or $d = 2$, respectively.
Its spectrum therefore consists of a single band $\omega_\vec{k} \in [0, 2^{d+1}J]$, and the on-site potential $\Delta$ constitutes a detuning with respect to the lower band edge. 

The Hamiltonian $H$ conserves the number of particles.
Hence, when studying the dynamics it produces for a given initial state with a well-defined number of particles, we can restrict the Hilbert space to the subspace of states with that same number of particles.
In the thermodynamic limit, we instead consider the filling fraction or, equivalently, the Fermi level $E_F$, which is defined in terms of the ground state of $H_B$ (which also conserves the number of particles) as the highest energy of the occupied bath eigenmodes.
This ground state is
\begin{equation}
\label{eq:Fermi sea}
    \ket{\rm FS} = \Biggl(\, \prod_{\omega_\vec{k} < E_F} b_\vec{k}^\dag\Biggr) \ket{\rm vac} \,,
\end{equation}
where ``FS'' stands for ``Fermi sea'', and $\ket{\rm vac}$ denotes the vacuum state.
A Fermi level $E_F \leq 0$ corresponds to an empty bath, whereas $E_F \geq 2^{d+1} J$ corresponds to a fully occupied bath.

The standard Hamiltonian employed in quantum optics to describe quantum emitters coupled to structured (photonic) environments is recovered from Eq.~\eqref{eq:Hamiltonian} by replacing the impurity and bath modes with spin and boson operators, respectively. This spin/boson model and our fermionic model share the same single-particle physics, however they differ in their many-body physics due to the different particle statistics and the intrinsic non-linearity introduced by spin impurities. We therefore want to move beyond the single-excitation regime and in the rest of this section we develop the formalism that makes this possible.

\subsection{Lattice Green's function}
Throughout this work, we will repeatedly refer to the lattice Green's function of the bath, which we also call the \emph{self-energy function} and which is defined in the thermodynamic limit ($L\to\infty$) as \cite{Economou2006}
\begin{equation}
\label{eq:self-energy}
    \Sigma(z, \vec{r}) = g^2 \int_\mathrm{BZ} \frac{d^d\vec{k}}{(2\pi)^d} \frac{e^{i\vec{k}\cdot\vec{r}}}{z - \omega_\vec{k}} \,.
\end{equation}
For a 1D bath, an analytical expression can be obtained (see Appendix~\ref{app:self-energy}). 
However, in 2D, there is no general analytical expression, except for $\vec{r} = (n, n)$ ($n \in \mathbb{Z}$), in which case $\Sigma(z,\vec{r})$ can be expressed in terms of hypergeometric functions~\cite{katsura_lattice_1971}.
The lattice Green's function does satisfy certain recursion relations that can in principle be used to compute it for other values of $\vec{r}$~\cite{morita_useful_1971, berciu_computing_2009}. 
In practice, however, we find it more convenient to reduce Eq.~\eqref{eq:self-energy} to a 1D integral which can then be computed numerically in a simple manner (see Appendix~\ref{app:self-energy}).

\subsection{Markovian master equation}

Under a Born-Markov approximation~\cite{BreuerPetruccione2007,schuetz_superradiance-like_2012}, the bath modes can be traced out (see Appendix~\ref{app:master_equation}) to obtain a master equation for the impurity density matrix $\rho$,
\begin{equation}
\label{eq:master_equation}
    \dot{\rho} = -i[\widetilde{H}_S, \rho] + \mathcal{D}_> \rho + \mathcal{D}_< \rho \,,
\end{equation}
where we have defined $\widetilde{H}_S = H_S + \sum_{m,n} J_{mn} c_m^\dag c_n$ and
\begin{subequations}
\begin{align}
    \mathcal{D}_> \rho & = \sum_{m,\, n} \Gamma^>_{mn} \left(c_n \rho c_m^\dag - \frac{1}{2} \{c_m^\dag c_n, \rho\} \right) \,, \\
    \mathcal{D}_< \rho & = \sum_{m,\, n} \Gamma^<_{mn} \left(c^\dag_m \rho c_n - \frac{1}{2} \{c_n c_m^\dag, \rho\} \right) \,.
\end{align}
\end{subequations}
Physically, $\widetilde{H}_S$ captures the coherent dynamics, including coherent interactions mediated by the bath, while $\mathcal{D}_>$ and $\mathcal{D}_<$ describe incoherent processes of fermion emission from the impurities into the bath and fermion absorption from the bath into the impurities, respectively. 

The coherent couplings $J_{mn}$ and collective absorption/emission rates $\Gamma_{mn}^\lessgtr$ are related to the self-energy function~\eqref{eq:self-energy} by $\Sigma(\Delta^\pm, \vec{r}_{nm}) = J_{mn} \mp i\Gamma_{mn}/2$ and
\begin{equation}
\label{eq:gamma_lessgtr}
    \Gamma_{mn}^\lessgtr = \begin{cases}
        \Gamma_{mn} \,, & \Delta \lessgtr E_F \\
        0 \,,           & \text{otherwise}
    \end{cases} \,,
\end{equation}
where $x^\pm \equiv x \pm i0^+$ and $\vec{r}_{nm} \equiv \vec{r}_m - \vec{r}_n$. 
Note how Eq.~\eqref{eq:gamma_lessgtr} implies that only one of the two dissipators, $\mathcal{D}_<$ or $\mathcal{D}_>$, is present for any given values of $\Delta$ and $E_F$. Qualitatively, the master equation dynamics are very intuitive: if $\Delta$ lies inside the band but above the Fermi level, there will be unoccupied bath modes available at the impurity energy, allowing occupied impurities to emit a fermion at that energy. 
Conversely, below the Fermi level, occupied bath modes resonant with the impurities will populate vacant impurities.
If $\Delta$ lies outside the band, $\Gamma_{mn}$ vanishes (essentially as a consequence of Fermi's Golden Rule) and the dynamics are purely coherent. 
Notably, the associated interaction strengths $J_{mn}$ are identical to those of conventional two-level systems coupled to a bosonic bath. 
This reflects the fact that the interactions are mediated by certain localised \emph{single-particle} eigenstates~\cite{leonforte_dressed_2021}, which are of course insensitive to the particle statistics.

The Markovian approximation underlying Eq.~\eqref{eq:master_equation} amounts to the assumption of a memory-less bath which remains approximately in thermal equilibrium throughout the dynamics (see Appendix~\ref{app:master_equation}). This assumption is valid only when $\rho(\omega)f(\omega)\approx\rho(\Delta)f(\Delta)$ for $\omega$ varied across an interval of width $\sim g^2/J$ around $\Delta$~\cite{schuetz_superradiance-like_2012}. Here, $\rho(\omega)=-\mathrm{Im}\Sigma(\omega^+,\vec{0})/(\pi g^2)$ is the density of states in the bath and $f(\omega)$ is the Fermi-Dirac distribution characterising the equilibrium state of the bath, which we assume to be the zero-temperature Fermi sea state~\eqref{eq:Fermi sea}, $f(\omega)=\Theta(E_F-\omega)$. This implies that Eq.~\eqref{eq:master_equation} will fail to accurately describe the dynamics when $\Delta$ lies close to $E_F$ (where $f(\omega)$ changes abruptly) or whenever $\rho(\omega)$ varies significantly around $\Delta$. In these regimes, we observe non-Markovian effects. To describe them, we develop an exact description of the system in the next two subsections.

\subsection{Free-fermion formalism}
The Hamiltonian~\eqref{eq:Hamiltonian} is quadratic in the fermionic operators, reflecting the fact that the fermions are \emph{non-interacting}. Below, we note some of the key properties of such free-fermion systems. \vspace{-1em}

\subsubsection{Fermionic Fock states}
Consider a set of $M$ independent fermionic creation and annihilation operators $\{\psi^\dag_l, \psi_l\}_{l=1}^M$. We assume the set to be complete in the sense that the states $\{\ket{\psi_l} \equiv \psi_l^\dag \ket{\rm vac}\}_{l=1}^M$ form an orthonormal basis of the single-particle Hilbert space. A basis of the whole many-body Hilbert space is then given by the Fock states
\begin{equation}
\label{eq:Fockstates}
    \ket{\vec{n}}\equiv \left(\psi^\dag_1\right)^{n_1}\dots\left(\psi^\dag_M\right)^{n_M}\ket{\rm vac}\,,
\end{equation}
where $\vec{n}\in\{0,1\}^M$, such that its $l$-th entry $n_l$ represents the occupation of the $l$-th mode in the state. It is straightforward to show that for an arbitrary quadratic (or \emph{one-body}) operator \footnote{In principle, terms of the form $\psi_l \psi_m$ and $\psi^\dag_l \psi^\dag_m$ are allowed in a completely general quadratic operator; however, in this work we focus only on quadratic operators conserving the total number of particles.} $O = \sum_{l, m} O_{lm} \psi^\dag_l \psi_m$,
\begin{multline}
    \bra{\bm n'} O \ket{\bm n} = \sum_{l,\, m} O_{lm} n'_l n_m \delta_{\bm{n}' - \bm{e}^l, \bm{n} - \bm{e}^m} \\
    \times (-1)^{\sum_{j < l} n'_j + \sum_{j < m} n_j} \,,
\end{multline}
where $\bm{e}^l$ is the unit vector with zeros everywhere except for the $l$-th component. 
Note that $O_{lm} = \bra{\psi_l} O \ket{\psi_m}$, and $\bra{\bm n'} O \ket{\bm n}\neq 0$ only if $\ket{\bm n}$ and $\ket{\bm n'}$ have the same number of particles and differ at most by one pair of occupation numbers. 
In particular, 
\begin{equation}
\label{eq:expectationvalue}
    \bra{\bm n} O \ket{\bm n} = \sum_l n_l O_{ll} \,.
\end{equation}
One natural choice of operator basis are the single-particle eigenmodes $\{\phi^\dag_l, \phi_l\}_{l=1}^M$, which diagonalise the Hamiltonian as $H=\sum_l \varepsilon_l \phi^\dag_l \phi_l$. 
The many-body eigenstates of $H$ are then given by Fock states of the form~\eqref{eq:Fockstates} in the basis formed by these eigenmodes. 
In particular, the ground state for a given Fermi level (filling) is
\begin{equation}
\label{eq:GS}
    \ket{\rm GS} = \Biggl(\, \prod_{\varepsilon_l < E_F} \phi_l^\dag \Biggr) \ket{\rm vac} \,.
\end{equation}
According to Eq.~\eqref{eq:expectationvalue}, ground-state expectation values of arbitrary one-body operators therefore simply read
\begin{equation}
\label{eq:expectationvalue_GS}
    \mean{O}_\mathrm{GS} = \sum_{\varepsilon_l < E_F} \bra{\phi_l} O \ket{\phi_l} \,.
\end{equation}

\subsubsection{Time-evolved expectation values}
The basis of single-particle eigenmodes $\{\phi^\dag_l, \phi_l\}_{l=1}^M$ also lends itself to describing time evolution: the expectation value of a generic quadratic operator $O$ at time $t$ is
\begin{equation}
\label{eq:quadraticevolution}
    \mean{O}_t = \sum_{l,\, m} O_{lm} e^{i(\epsilon_l - \epsilon_m)t} \mean{\phi^\dag_l \phi_m}_0 \,,
\end{equation}
where $\langle\,\cdot\,\rangle_0$ denotes an expectation value in the initial state. In studying the dynamics of our system, we will repeatedly consider the long-term average (LTA),
\begin{equation}
\label{eq:LTAdefinition}
    \overline{\mean{O}} 
    \equiv \lim_{T\to\infty}\int_0^T dt\, \frac{\mean{O}_t}{T}\,.
\end{equation}
It follows from Eq.~\eqref{eq:quadraticevolution} that for a quadratic operator,
\begin{equation}
\label{eq:LTA_quadratic}
    \overline{\mean{O}} = \sum_{l,\, m} O_{lm} \delta_{\epsilon_l,\epsilon_m} \mean{\phi^\dag_l \phi_m}_0 \,.
\end{equation}
i.e.\ it is determined entirely by the correlations between eigenmodes in the initial state. When $\mean{O}_t$ converges to a specific value 
as $t\to\infty$, this limit coincides with $\overline{\mean{O}}$.

\subsubsection{Gaussian state formalism}
The theory of quadratic Hamiltonians is inextricably linked with the theory of Gaussian states, since Gaussianity is preserved under evolution with a quadratic Hamiltonian. 
A remarkable property of Gaussian states is the fact that any even-order correlations can be expressed entirely in terms of 2-point correlations according to Wick’s Theorem.
Thus, Gaussian states are characterised fully by their covariance matrix \cite{surace_fermionic_2022,barthel_solving_2022,windt_local_2021,hackl_geometry_2020}. This property has been exploited to develop a number of methods for studying fermionic condensed matter models, such as a time-dependent formalism for calculating spectral functions~\cite{schonhammer_time-dependent_1978} and the \emph{Functional Determinant} approach to electron transport and dynamics~\cite{abanin_tunable_2004,levitov_electron_1996,d'Ambrumenil_Fermi_2005,benjamin_single-band_2014,schmidt_universal_2018}.

If the total number of particles in the system is well-defined and preserved during evolution, as is the case for the Hamiltonian~\eqref{eq:Hamiltonian}, it is sufficient to consider the correlation matrix $\bm C(t)$ (or equivalently the one-particle reduced density matrix) whose elements are $C_{lm}(t) = \mean{\psi^\dag_l \psi_m}_t$.
It evolves as~\cite{surace_fermionic_2022,barthel_solving_2022}
\begin{equation}
\label{eq:correlation_matrix}
    \bm{C}(t) = \bm{U}^\dag(t) \bm{C}(0) \bm{U}(t) \,,
\end{equation}
where $U_{lm}(t) = \bra{\psi_l} e^{-iH^Tt} \ket{\psi_m}$. 
If the system is time-reversal invariant, as in our case, then $H$ is real and $H^T = H$, so that $\bm{U}(t)$ is simply the matrix representation of the single-particle time-evolution operator. Crucially, the correlation matrix formalism not only provides us with an economical way of computing real-time dynamics of large systems, but also allows us to extend some of the analytical tools typically associated with the single-excitation regime to the many-body regime, as we discuss in the next section.

\subsection{Resolvent formalism}
The resolvent formalism allows the exact calculation of the transition amplitudes between eigenstates of a Hamiltonian $H_0$ induced by an interaction $V$ \cite{Cohen-Tannoudji1998}.
In the context of quantum emitters coupled to reservoirs, it has proven powerful for understanding the non-Markovian features of the dynamics in the single-excitation sector. 
\vspace{-.5em}

\subsubsection{Basic formalism}
In our case, $H_0 = H_S + H_B$ with single-particle eigenstates $\ket{e_n}\equiv c_n^\dagger\ket{\rm vac}$  and $\ket{\vec{k}}\equiv b_\vec{k}^\dagger\ket{\rm vac}$.
For now, we will focus on the case of a single impurity and re-label $\ket{e_1}\equiv\ket{e}$, although the formalism can also be applied to configurations with multiple impurities~\cite{gonzalez-tudela_markovian_2017}. 
The transition amplitudes $\mathcal{A}_e(t)\equiv\bra{e}e^{-iHt}\ket{e}$, $\mathcal{A}_\vec{k}(t)\equiv\bra{\vec{k}}e^{-iHt}\ket{e}$, and $\mathcal{A}_{\vec{q}\vec{k}}(t)\equiv\bra{\vec{q}}e^{-iHt}\ket{\vec{k}}$ can be computed as
\begin{equation}
\label{eq:resolvent_integral}
    \mathcal{A}_\alpha(t) = \frac{i}{2\pi} \int_{-\infty}^\infty dE\, G_\alpha(E^+) e^{-iEt} \,,
\end{equation}
with the propagators defined (with $\Sigma_e(z)\equiv\Sigma(z,\vec{0})$) as
\begin{subequations}
\label{eq:propagators}
\begin{align}
    & G_e(z) = \frac{1}{z - \Delta - \Sigma_e(z)}\,, \\
    & G_\vec{k}(z) = \frac{g}{\sqrt{L^d}} \frac{G_e(z)}{(z - \omega_\vec{k})}\,, \\
    & G_{\vec{q}\vec{k}}(z) = \frac{1}{z-\omega_\vec{k}}\left(\delta_{\vec{q}\vec{k}}+\frac{g}{\sqrt{L^d}}G_\vec{q}(z)\right)\,.
\end{align}
\end{subequations}
Formally, these are the matrix elements of the resolvent of the Hamiltonian between single-particle eigenstates of $H_0$~\cite{leonforte_dressed_2021}. The origin of non-Markovian dynamical features then lies in the analytic properties of the $G_\alpha(z)$~\cite{Cohen-Tannoudji1998, gonzalez-tudela_markovian_2017}.

\subsubsection{Multi-excitation quench dynamics}
We now show how the dynamics of a multi-excitation state can be expressed in terms the amplitudes $\mathcal{A}_\alpha(t)$, thus effectively reducing the many-body problem to a (well-studied) single-particle problem. 
Since we are working in the Gaussian regime, we can characterize  the state of the system completely by its 2-point correlations $\mean{c_n^\dagger c_m}$, $\mean{c_n^\dagger b_\vec{k}}$, and $\mean{b_\vec{k}^\dagger b_\vec{q}}$. 

Consider the initial state $\ket{\psi_0}=c^\dagger\ket{\rm FS}$, which will be relevant for the following discussion. For this state,
\begin{equation}
\label{eq:impurity_ocupation_resolvent}
\begin{split}
    \mean{c^\dagger c}_t
    &=\underbrace{\bra{e}e^{iHt}\ket{e}}_{\mathcal{A}^*_e(t)}\underbrace{\bra{e}e^{-iHt}\ket{e}}_{\mathcal{A}_e(t)} \\
    &\qquad\quad+\sum_{\omega_\vec{k}<E_F}\underbrace{\bra{\vec{k}}e^{iHt}\ket{e}}_{\mathcal{A}^*_\vec{k}(t)}\underbrace{\bra{e}e^{-iHt}\ket{\vec{k}}}_{\mathcal{A}_\vec{k}(t)} \,, \\
\end{split}
\end{equation}
where we have first re-written the correlator as $\mean{c^\dagger c}_t = \bra{\psi_0} e^{iHt} \, c^\dagger c \, e^{-iHt} \ket{\psi_0}$ and then applied Eq.~\eqref{eq:expectationvalue} to this expectation value, using the fact that $\ket{\psi_0}$ is a fermionic Fock state of the form~\eqref{eq:Fockstates}, and $H$ conserves the total number of particles. We have also used the fact that $\mathcal{A}_\alpha(-t)=\mathcal{A}_\alpha(t)$ for a time-reversal invariant model.
The other 2-point correlators can be obtained analogously  as
\begin{subequations}
\label{eq:correlations_resolvent}
\begin{align}
    & \mean{c^\dagger b_\vec{k}}_t = \mathcal{A}^*_e(t)\mathcal{A}_\vec{k}(t)+\sum_{\omega_\vec{q}<E_F}\mathcal{A}_\vec{q}(t)\mathcal{A}^*_{\vec{q}\vec{k}}(t)\,, \\
    & \mean{b_\vec{k}^\dagger b_\vec{q}}_t = \mathcal{A}^*_\vec{k}(t)\mathcal{A}_\vec{q}(t)+\sum_{\omega_\vec{p}<E_F}\mathcal{A}_{\vec{p}\vec{k}}(t)\mathcal{A}^*_{\vec{p}\vec{q}}(t)\,.
\end{align}
\end{subequations}
Together with Eq.~\eqref{eq:resolvent_integral}, this gives us a way to calculate the exact dynamics (semi-)analytically. 
More generally, the amplitudes computed via Eq.~\eqref{eq:resolvent_integral} provide an expression for $\bm{U}(t)$ in a particular basis, which can then be used to compute the time-dependence of the correlations for any initial state via Eq.~\eqref{eq:correlation_matrix}.

These formulae also provide a basis for developing various approximations.
For example, one can generalize the \emph{single-pole} or \emph{Wigner-Weisskopf approximation}~\cite{Cohen-Tannoudji1998,gonzalez-tudela_markovian_2017} to the many body regime, replacing $\Sigma_e(z)$ by $\Sigma_e(\Delta^+)$ in the different propagators, but still using Eqs.~\eqref{eq:impurity_ocupation_resolvent} and \eqref{eq:correlations_resolvent} to compute the time-dependence of the two-point correlations. 
This generally leads to a different dynamics than that predicted by the master equation, in contrast to what happens in the single-particle case, where both approaches lead to the same dynamics. This is because the conditions for the Wigner-Weisskopf and master equation descriptions to be accurate, i.e.\ essentially $\rho(\omega)\approx\rho(\Delta)$ and $\rho(\omega)f(\omega)\approx\rho(\Delta)f(\Delta)$, respectively, are not equivalent for a non-trivial bath particle distribution $f(\omega)$.

\subsubsection{Long-term averages}
We can also compute LTA expectation values easily from the integral expressions for the transition amplitudes.
By Eq.~\eqref{eq:LTAdefinition}, only the real poles of the propagators $G_\alpha(z)$, which contribute terms to the $\mathcal{A}_\alpha(t)$ that do not decacy in time, will contribute to the LTA of the correlation matrix elements.
From Eqs.~\eqref{eq:propagators}, we see that $G_e(z)$ has two real poles, $z = \omega_\pm$, which are solutions to
\begin{equation}
\label{eq:pole_equation}
    z - \Delta - \Sigma_e(z) = 0 \,,
\end{equation}
while $G_\vec{k}(z)$ shares the real poles of $G_e(z)$ and has the additional real pole $\omega_\vec{k}$. 
Similarly, $G_{\vec{q}\vec{k}}(z)$ shares all poles of $G_\vec{k}(z)$ and has the additional pole $\omega_\vec{q}$. 
Thus, the LTA impurity occupation, for instance, can be computed as
\begin{equation}
\label{eq:LTA_population}
\begin{split}
    \overline{\mean{c^\dagger c}}
    &=\sum_{z=\omega_\pm}R_e(z)^2\left(1+\int_{\omega_\vec{k}<E_F}\frac{d^d\vec{k}}{(2\pi)^d}\frac{g^2}{(z-\omega_\vec{k})^2}\right) \\
    &\qquad+g^2\int_{\omega_\vec{k}<E_F}\frac{d^d\vec{k}}{(2\pi)^d}\abs{G_e(\omega_\vec{k}^+)}^2 \,,
\end{split}
\end{equation}
where the first and second terms are due to the real poles $\omega_\pm$ of $G_e(z)$ and the additional pole $\omega_\vec{k}$ of $G_\vec{k}(z)$, respectively, and $R_e(z)=[z-\partial_z\Sigma_e(z)]^{-1}$ is the residue associated with simple poles of $G_e(z)$.

\subsection{Fermions vs. bosons}
\label{sec:fermions_vs_bosons}
Before proceeding, it is worth noting that the theory we have developed is also largely applicable to the bosonic counterpart of our model, where the impurity and bath modes describe non-interacting bosons (in which case the Hamiltonian is again quadratic). 

Firstly, Eq.~\eqref{eq:expectationvalue} is formally identical in the bosonic case. In fact, since the single-particle eigenspectrum of quadratic models is not sensitive to particle statistics, the dynamics generated by our Hamiltonian according to Eq.~\eqref{eq:correlation_matrix} are also identical. From the perspective of Gaussian states, this is due to the fact that the Hamiltonian preserves occupation number and thus only generates the subset of Gaussian transformations common to both the fermionic and bosonic Gaussian state families~\cite{hackl_geometry_2020,windt_local_2021}.

The crucial difference between fermions and bosons therefore lies in the possible initial states. 
It is the choice of a Fermi sea state in the bath---which is not exclusively fermionic \emph{per se}, but which of course does not describe the ground state of a bosonic bath---which leads to the explicit form of Eqs.~\eqref{eq:master_equation},~\eqref{eq:expectationvalue_GS}, and~\eqref{eq:impurity_ocupation_resolvent}-\eqref{eq:correlations_resolvent}. Indeed, general bosonic Fock states are not Gaussian, and therefore, while Eq.~\eqref{eq:correlation_matrix} still captures the evolution of the bosonic two-point correlations for the intial states we consider in this work, these do not in general characterise the full state.
The transition amplitudes $\mathcal{A}_\alpha(t)$ themselves are once again independent of particle statistics, and so the notion of describing the many-body physics of the model in terms of its single-particle properties translates straightforwardly to the bosonic regime. Indeed, there also exists a bosonic version of Functional Determinant formalism mentioned above~\cite{Nazarov2003,xi_supercooling_2016}.

\section{Single impurity}
\label{sec:singleimpurity}
Having developed an extensive theoretical toolbox to study the Hamiltonian~\eqref{eq:Hamiltonian}, we now apply it first to the case of a single impurity. Without loss of generality, we assume that the impurity is coupled to the bath at $\vec{r}_1\equiv\vec{0}$, so that the Hamiltonian becomes
\begin{equation}
\label{eq:Hamiltonian_RLM}
    H=\Delta c^\dag c+\sum_\vec{k}\omega_\vec{k}b_\vec{k}^\dag b_\vec{k}+\frac{g}{\sqrt{L^d}}\sum_\vec{k}\left(c^\dagger b_\vec{k}+ \mathrm{H.c.}\right) \,.
\end{equation}
From a condensed-matter point of view, this Hamiltonian constitutes a minimal model for certain mesoscopic systems (semiconductor structures, e.g.\ quantum dots)~\cite{Haug2008}. Moreover, it represents a limiting case of more complicated impurity models which include spin degrees of freedom and / or interactions between fermions: the \emph{Interacting Resonant Level Model}~\cite{borda_theory_2007}, the \emph{Single-impurity Anderson Model}~\cite{anderson_localized_1961}, and the \emph{Kondo Model}~\cite{anderson_exact_1970,toulouse_infinite_1970}. As noted above, from the perspective of quantum optics, it shares the same single-particle physics as the spin/boson models that have been studied recently (see e.g.\ Ref.~\cite{gonzalez-tudela_markovian_2017}). However, the fermionic character of the excitations leads to decidedly different many-body physics, as we demonstrate in this section.

The simplicity of our model allows us to exploit powerful techniques from quantum optics to describe the system exactly, in an unprecedented level of detail. Specifically, we firstly analyze the ground state properties of the Hamiltonian~\eqref{eq:Hamiltonian_RLM} in detail. Contrary to the standard approach followed when studying the RLM, we account for the full structure of the (finite) band of bath modes, rather than linearising around the Fermi level. Thereby, we are able to derive formally exact expressions for the ground-state correlations, allowing us to generalise existing results on the impurity-bath correlations in 1D~\cite{ghosh_real-space_2014} and to thoroughly examine these correlations in 2D.

Secondly, we consider the dynamics of the model, for an initial state in which the impurity mode is occupied and the bath is in the Fermi sea state~\eqref{eq:Fermi sea}. The formalism developed in the previous section allows us to study these dynamics in the non-Markovian regime exactly and without any approximations.

\subsection{Ground-state features}

\subsubsection{Single-particle eigenstates}
The single-particle spectrum of $H$ comprises two distinct types of eigenstates: a continuum of scattering eigenstates $\ket{\phi_\vec{k}}$, parametrised by quasi-momenta $\vec{k}\in {\rm BZ}$, with energies $\omega_\vec{k}$ \cite{zhou_controllable_2008,leonforte_dressed_2021},
\begin{equation}
    \label{eq:scattering_state}   
    \ket{\phi_\vec{k}}=\frac{\abs{G_e(\omega_\vec{k}^+)}}{\sqrt{L^d}}\left\{g\ket{e}+\sum_j\left[\frac{e^{i\vec{k}\cdot\vec{r}_j}}{G_e(\omega_\vec{k}^+)}+\Sigma\left(\omega_\vec{k}^+,\vec{r}_j\right)\right]\ket{j}\right\}\,,
\end{equation}
where $\ket{j}\equiv b^\dag_j \ket{\rm vac}$, and two bound states (BS) $\ket{\phi_\pm}$, with energies $\omega_\pm$ given by the real solutions to the pole equation~\eqref{eq:pole_equation}, satisfying $\omega_+>2^{d+1}J$ and $\omega_-<0$~\cite{calajo_atom-field_2016},
\begin{equation}
    \ket{\phi_\pm}=\sqrt{R_e(\omega_\pm)}\left\{\ket{e}+\frac{1}{g}\sum_j\Sigma\left(\omega_\pm,\vec{r}_j\right)\ket{j}\right\}\,.
\end{equation}
The presence of two bound states is due to the divergent Lamb shift (with opposite sign) at the two band edges and is therefore a consequence of the finite bandwidth of the bath. In fact, for a finite but sufficiently large bath, coupling to the impurity mode results in two localized modes and $L-1$ scattering eigenstates in the band range.

The many-body ground state is a product state of the form shown in Eq.~\eqref{eq:GS}, in which these single-particle eigenstates of energies up to $E_F$ are occupied. In the following, we focus on the case where the Fermi level lies within the band, $0\leq E_F \leq 2^{d+1} J$, such that only the lower bound state (LBS) and the scattering eigenstates of energies $\omega_\vec{k} < E_F$ are occupied in the ground state.

\subsubsection{Impurity occupation}
Using the single-particle eigenstate formulae above, the ground-state impurity occupation in the thermodynamic limit takes the form
\begin{equation}
\label{eq:GS_population}
    \mean{c^\dagger c}_\mathrm{GS}=R_e(\omega_-)+g^2\int_{\omega_\vec{k}<E_F}\frac{d^d\vec{k}}{(2\pi)^d}\abs{G_e(\omega_\vec{k}^+)}^2\,.
\end{equation}
If the coupling is small enough and the detuning is deep enough in the band, we can neglect the first term, which is the contribution from the LBS~\cite{gonzalez-tudela_markovian_2017}, since $R_e(\omega_-)=\langle\phi_-|c^\dagger c|\phi_-\rangle$ is small.
Then, when $\Delta \approx E_F$, we can approximate 
$\Sigma_e(\omega^+) \approx \Sigma_e(E_F^+) = \delta\omega_F - i \Gamma_F/2$, so that
\begin{equation}
\label{eq:GS_population_wide-band}
    \mean{c^\dag c}_{\rm GS} \approx \frac{\Gamma_F}{2\pi} \int_0^{E_F} \frac{d\omega}{(\omega - \Delta- \delta\omega_F)^2 + (\Gamma_F/2)^2} \,.
\end{equation}
If we additionally extend the lower integration limit to $-\infty$ under an infinite-band approximation \cite{parthenios_transient_2021},
\begin{equation}
\label{eq:Keldysh_population}
    \mean{c^\dag c}_{\rm GS} \approx \frac{1}{2} + \frac{1}{\pi} \tan^{-1}\left[\frac{2}{\Gamma_F}(E_F - \Delta - \delta\omega_F)\right] \,.
\end{equation}
Thus, the impurity is either occupied or empty for $\Delta \ll E_F$ or $\Delta \gg E_F$, respectively. 
The transition point occurs in a range of energies of width $\sim \Gamma_F$ around $E_F-\delta\omega_F$. 
Clearly, this reasoning breaks down in the 2D bath at the Van Hove singularity ($E_F \approx 4J$), where $\Gamma_F$ diverges~\cite{Economou2006}. 
In this case, a simplified integral expression for $\mean{c^\dag c}_{\rm GS}$ can be obtained by substituting an asymptotic expression for $\Sigma_e(\omega^+)$ as $\omega\to 4J$. 
However, this does not lead to a qualitative change in the dependence of $\mean{c^\dag c}_{\rm GS}$ on $\Delta$.

\subsubsection{Impurity screening cloud}

A more salient ground-state property of the single-impurity model is an emergent length scale associated with an \emph{impurity screening cloud}, which constitutes a hallmark feature of generic impurity models~\cite{Hewson1993}. 
It is reflected, for example, in the spatial dependence of the envelope of $\mean{c^\dagger b_j}_\mathrm{GS}$~\cite{borda_kondo_2007,ghosh_real-space_2014,ghosh_dynamics_2015}, which we denote here by $F(\vec{r}_j)$. Using the expressions for the single-particle eigenstates and Eq.~\eqref{eq:expectationvalue_GS}, in the thermodynamic limit,
\begin{multline}
\label{eq:GS_cloud}
    \mean{c^\dag b_j}_{\rm GS} = \frac{R_e(\omega_-)}{g} \Sigma(\omega_-, \vec{r}_j) \\
    + g \int\limits_{\mathclap{\omega_\vec{k} < E_F}} \frac{d^d\vec{k}}{(2\pi)^d} \, \left[G_e(\omega_\vec{k}^-) e^{i \vec{k} \cdot \vec{r}_j} + \abs{G_e(\omega_\vec{k}^+)}^2 \Sigma(\omega_\vec{k}^+, \vec{r}_j)\right] \,.
\end{multline}
The first term, coming from the LBS, decays exponentially or faster as the distance $\abs{\vec{r}_j}$ increases, in the form $\sim \abs{\vec{r}_j}^{-(d-1)/2} e^{-\abs{\vec{r}_j}/\xi}$ \cite{michta2022}, and, as noted above, its prefactor $R_e(\omega_-)$ is very small if $\Delta$ lies sufficiently deep inside the band. 
The second term gives the contribution of the scattering eigenstates, and it decays algebraically, as $\sim \abs{\vec{r}_j}^{-\alpha}$, so it dominates at large distances. 
Previous works seem to suggest a constant exponent $\alpha = 1$ for $1\leq d\leq 3$~\cite{ghosh_real-space_2014}.
However, as we show in Appendix~\ref{app:screening}, this algebraic decay is the same as the one of the bath-bath correlations $\mean{b^\dagger_0 b_j}_{\rm FS}$ in an impurity-free bath with the same Fermi level, and is ultimately determined by the dimensionality, and the smoothness of the Fermi surface. 
For a smooth Fermi surface, $\alpha = (d + 1)/2$~\cite{Wen2007}.
Note that this algebraic dependence is expected for critical (gapless) phases.

\begin{figure}[!t]
\centering
\includegraphics{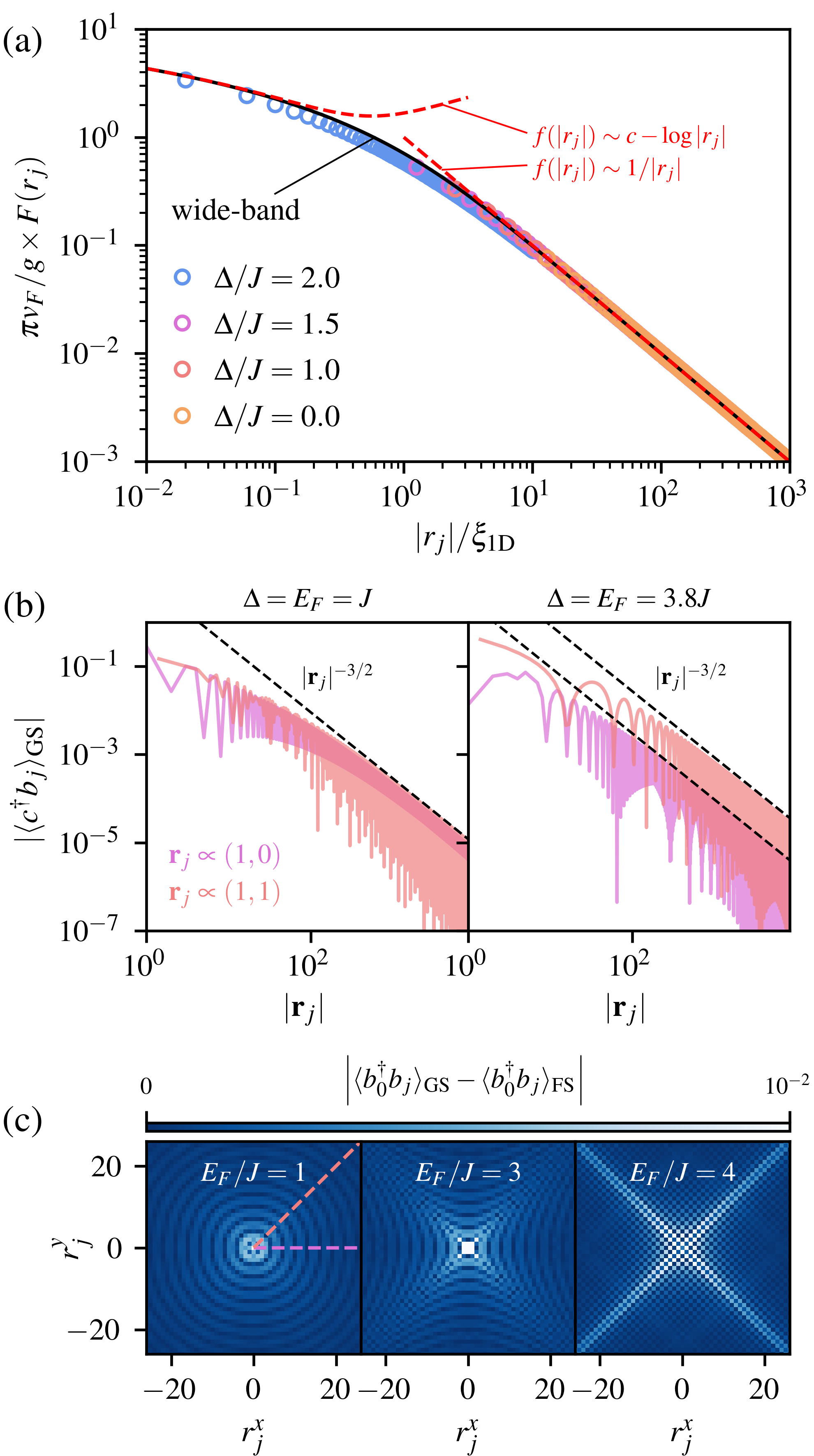}
\caption{\textsl{Single-impurity ground-state correlations.} a) Scaling of the 1D ground-state impurity-bath correlation envelope at fixed $E_F=2J$ and $g=0.2J$ for varied $\Delta$, computed using the wide-band formula~\eqref{eq:GS_cloud_wide-band} (black solid line) and by numerical integration of Eq.~\eqref{eq:GS_cloud} (coloured markers). The short- and long-range scalings~\eqref{eq:1D_GS_scalings} are also shown (red dashed lines). b) 2D ground-state impurity-bath correlation for two different $\Delta=E_F$ and $g=0.2J$, computed along two different directions as described in Appendix~\ref{app:screening}. For reference, two directions are indicated in the left pane of c). We also highlight the algebraic scaling of correlations. c) Deformation of the ground-state bath-bath correlations as given by Eq.~\eqref{eq:GS_bath_correlations}, for $g=0.2J$ and various $E_F=\Delta$.}
\label{fig:single-impurity_correlations}
\end{figure}

In 1D, if, in addition, the Fermi level is deep inside the band, we can make a wide-band approximation (see Appendix~\ref{app:screening}) and neglect the LBS contribution to obtain
\begin{equation}
\label{eq:GS_cloud_wide-band}
    \mean{c^\dag b_j}_{\rm GS} \approx \frac{g}{\pi v_F} \re\left[e^{ik_F\abs{r_j}}f\left(\frac{\abs{r_j}}{\xi_{\rm 1D}}\right)\right]\,,
\end{equation}
with $f(x) = -e^{ix} E_1(ix)$, where $E_1$ is the exponential integral function, $v_F = 2 g^2/\Gamma_F$ is the group velocity at the Fermi level, and
\begin{equation}
\label{eq:1D_screening_length}
    \xi_\mathrm{1D} = \left[\left(\frac{\Delta - E_F}{v_F}\right)^2 + \left(\frac{g}{v_F}\right)^4\right]^{-1/2} \,.
\end{equation}
We plot the envelope $F(r_j)$ of the approximate correlations in Fig.~\ref{fig:single-impurity_correlations}a, together with a numerical evaluation of the exact formula~\eqref{eq:GS_cloud}. 
Remarkably, good agreement between the two is still observed when $\Delta$ lies close to the band edge. The short- and long-range scaling of the correlations, also indicated in Fig.~\ref{fig:single-impurity_correlations}a, are dictated by
\begin{align}
\label{eq:1D_GS_scalings}
    f(x \ll 1) & \sim -\gamma - \log(x) - i\pi/2 \,, \\
    f(x \gg 1) & \sim - 1/x \,.
\end{align}
This generalises what was already observed in Ref.~\cite{ghosh_real-space_2014} for the special case $\Delta=E_F$: there is a transition from logarithmic to algebraic decay of the ground-state impurity-bath correlator at $\abs{r_j} \approx \xi_\mathrm{1D}$. 
In this sense, $\xi_{\rm 1D}$ can be interpreted as the size of the screening cloud. 

Owing to the lack of an analytical expression for $\Sigma(\omega_\vec{k}, \vec{r}_j)$, we cannot obtain a ``universal'' formula, as Eq.~\eqref{eq:GS_cloud_wide-band}, for the correlations in the 2D case.
Nonetheless, we compute them using an efficient numerical method (detailed in Appendix~\ref{app:screening}) that allows us to reach very large system sizes, and thus to observe the onset of their algebraic decay and determine the screening length for different directions, see Fig.~\ref{fig:single-impurity_correlations}b.
In this way, we can map out the shape of the screening cloud in 2D. 
One might expect certain degree of anisotropy, especially if the Fermi surface is highly anisotropic, for example, when the Fermi level is tuned close to the Van Hove singularity. 
However, for the cases considered, the screening length seems to have a rather small angular dependence, irrespective of the value of the Fermi level. 
If we instead look at the actual values of the correlations along different directions, we do find larger differences the more anisotropic the Fermi surface is.

A similar angular dependence can be observed in the ground-state bath-bath correlator $\mean{b^\dag_0 b_j}_\mathrm{GS}$, which in the thermodynamic limit reads, neglecting an exponentially decaying contribution from the LBS,
\begin{multline}
\label{eq:GS_bath_correlations}
    \mean{b^\dag_0 b_j}_\mathrm{GS} \approx \mean{b^\dag_0 b_j}_\mathrm{FS}+\int\limits_{\mathclap{\omega_\vec{k} < E_F}} \frac{d^2\vec{k}}{(2\pi)^2} \, \abs{G_e(\omega_\vec{k}^+)}^2\\\times\left[\frac{\Sigma_e(\omega_\vec{k}^-)}{G_e(\omega_\vec{k}^+)}\,e^{i\vec{k}\cdot\vec{r}_j}+(\omega_\vec{k}-\Delta)\Sigma(\omega_\vec{k}^+,\vec{r}_j)\right]\,.
\end{multline}
Here, we have defined $\mean{\,\cdot\,}_\mathrm{FS} = \bra{\mathrm{ FS}}\,\cdot\,\ket{\mathrm{FS}}$, i.e.\ the first term captures the ground-state correlations in a bath without an impurity, while the second term represents a local deformation of the correlations due to the impurity. We plot this deformation Fig.~\ref{fig:single-impurity_correlations}c.

\subsection{Quench dynamics}
We now consider the dynamics of the initial state $\ket{\psi_0} = c^\dag \ket{\rm FS}$. 
This amounts to initialising the bath in its ground state at the desired filling, populating the impurity, and turning on a coupling $g>0$ instantaneously at time $t=0$.
Below we describe the transient dynamics of both the impurity and the bath, as well as the steady state after the quench.

\subsubsection{Impurity occupation}
The dynamics of an initially empty bath and an excited impurity mode have been studied extensively in the spin/bosonic setting \cite{gonzalez-tudela_markovian_2017} and the results can be directly extrapolated to our model.
In this regime, for $\Delta$ sufficiently deep in the band of the bath, the impurity occupation displays approximately exponential decay at rate $\Gamma_0= - 2 \im \Sigma_e(\Delta^+)$, however when $\Delta$ lies close to the band edges, we observe fractional decay due to trapping of the excitation in the bound states~\cite{john_quantum_1990,john_spontaneous_1994,gonzalez-tudela_markovian_2017}.
Additionally, in the 2D bath, when $\Delta = 4J$ (i.e.\ tuned to the Van Hove singularity) we observe algebraic decay~\cite{gonzalez-tudela_markovian_2017,gonzalez-tudela_quantum_2017}.

\begin{figure}[!htb]
    \centering
    \includegraphics{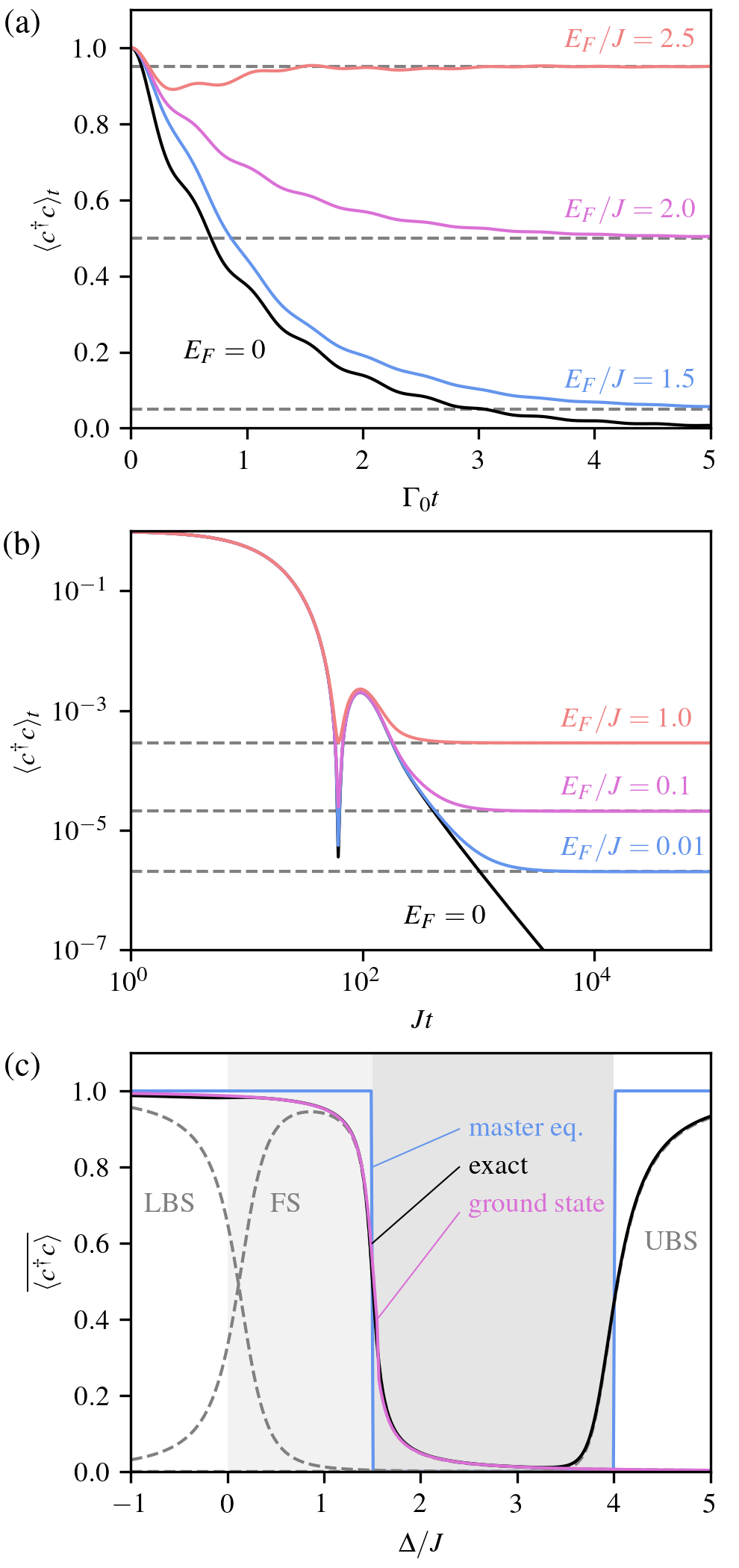}
    \caption{\textsl{Single-impurity quench dynamics.} a) Impurity occupation dynamics for a 1D bath with $\Delta=2J$ and $g=0.4J$ for various $E_F$ (solid) and corresponding LTA population~\eqref{eq:LTA_population} (dashed grey). b) Same as a), for a 2D bath at $\Delta=4J$ and $g=0.2J$. c) 1D LTA impurity occupation at $g=0.4J$ and $E_F=1.5J$ as a function of $\Delta$. We plot the exact result~\eqref{eq:single-impurity_LTA} (black) and the master equation prediction (blue), as well as the ground-state impurity occupation~\eqref{eq:GS_population} (purple). Dashed lines indicate contributions from the lower (LBS) and upper (UBS) bound states and the scattering states (FS). Grey shaded areas indicate the regions $0<\Delta<E_F$ (light) and $E_F<\Delta<4J$ (dark).}
    \label{fig:single-impurity_decay}
\end{figure}

For a non-zero filling, $E_F > 0$, the dynamics depend primarily on the value of the detuning relative to the Fermi level.
The master equation~\eqref{eq:master_equation} predicts $\mean{c^\dagger c}_t = e^{-\Gamma_0 t}$, when $\Delta > E_F$, and $\mean{c^\dagger c}_t = 1$, otherwise. 
However, the exact dynamics reveal some non-Markovian features. 
In addition to the effect of the bound states, we can observe fractional decay when $\Delta \approx E_F$, see Fig.~\ref{fig:single-impurity_decay}a. 
In 2D, when $\Delta = 4J$ we find that already at a very low value of $E_F$, the non-zero steady-state population induced by the Fermi level overwhelms the algebraic decay (see Fig.~\ref{fig:single-impurity_decay}b). 

If the coupling $g$ is sufficiently small, the impurity occupation converges to a specific value as time increases, which is given by its LTA $\overline{\mean{c^\dag c}}$, Eq.~\eqref{eq:LTA_population}.
This quantity is shown as a function of the detuning in Fig.~\ref{fig:single-impurity_decay}c. 
Notably, the profile of the LTA impurity occupancy is ``smeared out'' with respect to the master equation prediction around the band edges and the Fermi level, reflecting the non-Markovian possibility of decay or absence of decay even where Fermi's Golden Rule would prohibit it. Essentially, the observation of fractional decay around the Fermi level can therefore be viewed as a non-Markovian version of Pauli-inhibited spontaneous decay~\cite{busch_inhibition_1998}. In fact, comparing the expression for $\overline{\mean{c^\dag c}}$, Eq.~\eqref{eq:LTA_population}, and the impurity occupation in the ground state $\mean{c^\dag c}_{\rm GS}$, Eq.~\eqref{eq:GS_population}, one realizes that both are the same, except for contributions attributed to the bound states. As we show in the next section, this kind of relationship between the long-term average (steady state) expectation value and the ground-state expectation value actually holds for arbitrary one-body operators.

Before proceeding, it is worth noting that, for detunings deep in the band (where $R_e(\omega_\pm)$ is negligible) and in the weak coupling regime, the fractional decay around the Fermi level is described by Eq.~\eqref{eq:Keldysh_population}.
Notably, this formula has been previously obtained in the 1D case as the steady-state population for the Resonant Level Model in the Keldysh formalism~\cite{parthenios_transient_2021,anders_real-time_2005}. Of course, the result we have obtained in Eq.~\eqref{eq:LTA_population} is more general, since it does not rely on a weak-coupling approximation and incorporates the effects of the band edges.

\begin{figure*}[t]
\centering
\includegraphics{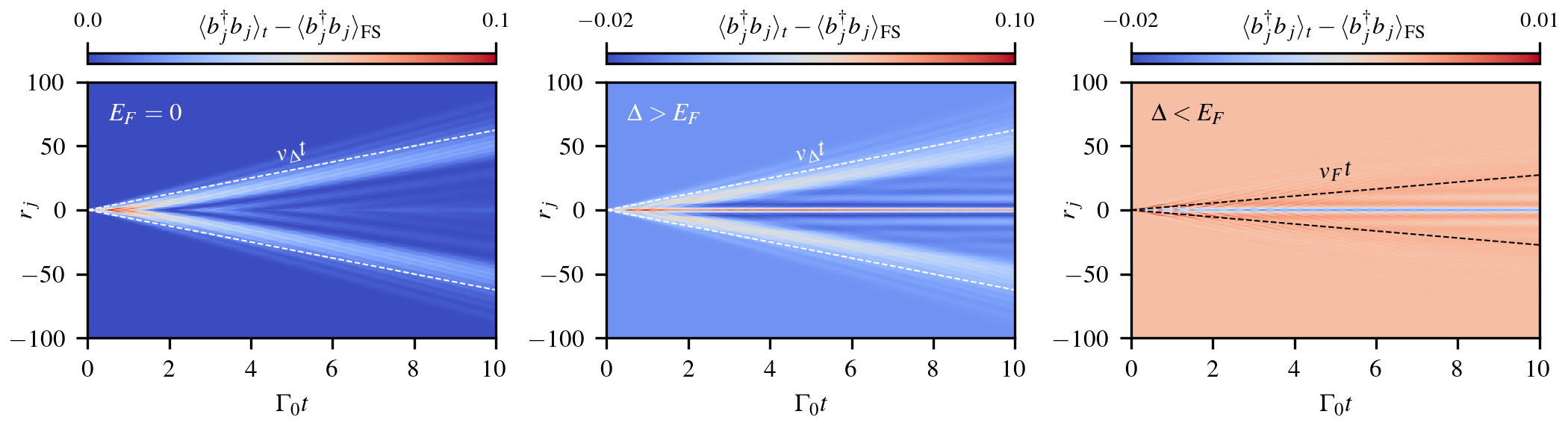}
\caption{\textsl{Propagation of excitations in a 1D bath.} 
Evolution of the excitation probability $\mean{b_j^\dag b_j}_t$ (arb. units) during the decay of a single impurity at fixed $k_\Delta=\pi/4$ and $g/J=0.4$ for an initially empty bath (left) and for $k_F=\pi/10$ (middle) and $k_F=9\pi/10$ (right), corresponding with $\Delta>E_F$ and $\Delta<E_F$, respectively.}
\label{fig:single-impurity_bath_excitation}
\end{figure*}

\subsubsection{Relaxation to the ground state}
Having established the relation between the steady-state and ground state impurity occupation above, we now discuss this relation for a general on-body operator $O$. According to Eq.~\eqref{eq:LTA_quadratic}, the LTA expectation value of any such operator after the quench is given by
\begin{equation}
\label{eq:single-impurity_LTA}
\begin{split}
    \overline{\mean{O}}
    &=\sum_{\nu=\pm}\mean{\phi_\nu^\dagger\phi_\nu}_0\bra{\phi_\nu}O\ket{\phi_\nu} \\
    &\qquad+\sum_{\vec{k},\vec{q}}\delta_{\omega_\vec{k}\omega_\vec{q}}\bra{\phi_\vec{k}}O\ket{\phi_\vec{q}}\mean{\phi_\vec{k}^\dagger\phi_\vec{q}}_0\,.
\end{split} 
\end{equation}
However, as we show in Appendix~\ref{app:thermodynamic_LTA}, in the thermodynamic limit we find that for an initial state of the form $\ket{\psi_0} = \left(c^\dag\right)^{n_e} \ket{\rm FS}$, with $n_e \in \{0, 1\}$, correlations between degenerate scattering eigenmodes are the same in the Fermi sea and in the ground state of the impurity model,
\begin{equation}
\label{eq:thermodynamic_LTA_result}
    \mean{\phi^\dag_\vec{k} \phi_\vec{q}}_0 = \delta_{\vec{q}\vec{k}} \Theta(E_F - \omega_\vec{k}) = \mean{\phi^\dag_\vec{k} \phi_\vec{q}}_{\rm GS} \,,
\end{equation}
if $\omega_\vec{k}=\omega_\vec{q}$. 
This result is physically intuitive: for a free bath mode, the impurities constitute a potential with a finite range, whose only effect is to induce a phase shift as the particle is scattered \cite{Taylor1972}. In terms of the LTA impurity occupation, Eq.~\eqref{eq:thermodynamic_LTA_result} implies that, in the thermodynamic limit,
\begin{equation}
\label{eq:LTAvsGS}
\begin{split}
    \overline{\mean{O}} = \mean{O}_\mathrm{GS}
    & + \bra{\phi_-} O \ket{\phi_-} \left(\mean{\phi^\dag_- \phi_-}_0 - 1\right) \\
    & + \bra{\phi_+} O \ket{\phi_+} \mean{\phi^\dag_+ \phi_+}_0 \,.
\end{split}
\end{equation}
For example, using Eq.~\eqref{eq:expectationvalue}, we have, for $\ket{\psi_0} = c^\dag \ket{\rm FS}$,
\begin{equation}
\label{eq:initial_BSoccupation}
    \mean{\phi_\pm^\dag \phi_\pm}_0 = R_e(\omega_\pm) \left(1 + \int_{\omega_\vec{k} < E_F} \frac{d^d\vec{k}}{(2\pi)^d}\frac{g^2}{(\omega_\pm-\omega_\vec{k})^2}\right) \,,
\end{equation}
and noting that $\bra{\phi_\pm} c^\dag c\ket{\phi_\pm} = R_e(\omega_\pm)$ 
we can check that Eq.~\eqref{eq:LTA_population} is consistent with Eq.~\eqref{eq:LTAvsGS}. 
In essence, Eq.~\eqref{eq:LTAvsGS} formalizes the idea that the initial state $\ket{\psi_0} = c^\dagger\ket{\rm FS}$ indeed relaxes to the ground state, except for a contribution from the BSs, which constitutes the effect of the finite bandwidth.

\subsubsection{Bath excitation dynamics}
Another interesting feature of the quench dynamics is the propagation of excitations through the bath following the decay of the impurity. This is captured by the time-dependent expecation value $\mean{b_j^\dag b_j}_t$. From Eqs.~\eqref{eq:correlations_resolvent},
\begin{equation}
\begin{split}
    \mean{b_j^\dag b_j}_t
    &=\left|\frac{1}{\sqrt{L^d}}\sum_\vec{k}\mathcal{A}_\vec{k}(t)\,e^{i\vec{k}\cdot\vec{r}_j}\right|^2 \\
    &\quad +\sum_{\omega_\vec{q}<E_F}\left|\frac{1}{\sqrt{L^d}}\sum_\vec{k}\mathcal{A}_{\vec{q}\vec{k}}(t)\,e^{i\vec{k}\cdot\vec{r}_j}\right|^2\,.
\end{split}
\end{equation}
We focus on the case where $\Delta$ lies in the band, in which case we observe an emission of matter-waves into the bath. 
We also restrict our attention to the case of a 1D bath, which allows us to obtain some analytical results. In particular, we can consider the limit $t\to\infty$ and $x/t\to\kappa$. Under a stationary phase approximation \cite{Bender1999},
\begin{equation}
\begin{split}
    &\mean{b_j^\dag b_j}_t
    \lesssim \mean{b_j^\dagger b_j}_\mathrm{GS} \\
    &\quad+\frac{1}{2\pi t\omega_\kappa}\left[\frac{\Theta(k_\kappa-k_F)}{C_\kappa^-}+\frac{\Theta(\pi-k_\kappa-k_F)}{C_\kappa^+}\right]^2\,,
\end{split}
\end{equation}
where $C_\kappa^\pm=\sqrt{(2J\pm\omega_\kappa-\Delta)^2+(g^2/\kappa)^2}$, $\omega_\kappa = \sqrt{(2J)^2-\kappa^2}$, and $k_\kappa =\arcsin(\kappa/(2J))$. 
From this expression, we can distinguish three distinct regimes (see Fig.~\ref{fig:single-impurity_bath_excitation}): for $E_F=0$, a wavefront forms and propagates at velocity $\sim v_\Delta$ (i.e.\ the group velocity at the detuning) due to the fact that the momenta of the emitted matter-waves are strongly peaked around the detuning momentum $k_\Delta=\arccos(1-\Delta/(2J))$ (with $\omega_{k_\Delta}=\Delta$)~\cite{sanchez-burillo_dynamical_2017}. For $0<E_F<\Delta$, the dynamics display qualitatively similar features to the case $E_F=0$, however now there is a significant occupation probability for sites around the impurity site, given by the ground state occupations. For $E_F>\Delta$, the emitter matter-waves are now predominantly confined to a new light cone, defined not by $v_\Delta$ but rather by the Fermi velocity $v_F$, because modes with $|k|<k_F$ are no longer available for propagation. \vspace{-.5em}

\section{Multiple impurities}
\label{sec:multipleimpurities}

We now turn to the problem of $N \geq 2$ impurities. 
We will focus our attention on the collective decay of a ``fully inverted'' state, i.e.\ the dynamics of the total impurity occupation $N_e=\sum_n c_n^\dag c_n$ starting from the initial state $\ket{\psi_0} = \left(\prod_n c_n^\dag\right) \ket{\rm FS}$. In the case of two-level impurities decaying collectively into an empty bath, this setting gives rise to a characteristic delayed \emph{superradiant burst} of emission, due to a build-up of coherence between the impurities~\cite{cardenas-lopez_many-body_2023}. While this setting of collective dynamics of multiple impurities coupled to a common bath is natural from the perspective of (matter-wave) quantum optics, it has, to our knowledge, not been studied in the context of condensed matter physics. In this section, we therefore discuss both the Markovian description of the dynamics as well as non-Markovian effects. We find that much of the conceptual discussion of the single-impurity dynamics can be intuitively extended to this more general case. In particular, we can again relate the steady-state of the system to its ground state and single-particle bound states in the spectrum of our model.

\subsection{General description}
In the Markovian regime, the collective decay is described by the master equation~\eqref{eq:master_equation}.  
The impurity correlations $C_{mn}(t) \equiv \mean{c^\dag_m c_n}_t$ thus evolve in the same way as for any quadratic model, as shown in Eq.~\eqref{eq:correlation_matrix}, but with a non-Hermitian effective Hamiltonian, which in the case $\Delta > E_F$ is given by $H_{\rm eff} = \sum_{m, n} \Sigma_{mn}(\Delta^+)\, c^\dag_m c_n$, where we have defined the \emph{self-energy matrix} $\bm{\Sigma}(z)$ as the matrix with elements $\Sigma_{mn}(z) \equiv\Sigma(z, \vec{r}_{mn})$.
Thus, for a fully inverted state, $\vec{C}(0) = I_{N\times N}$, the total impurity occupation $\mean{N_e}_t=\tr\bm{C}(t)$ evolves as 
\begin{equation}
\label{eq:collective_decay_markov}
    \mean{N_e}_t =\norm{e^{-i \bm{\Sigma}(\Delta^+) t}}_{\rm HS}^2\,,
\end{equation}
where $\norm{\,\cdot\,}_{\rm HS}$ denotes the Hilbert-Schmidt norm. With reference to the discussion in Sec.~\ref{sec:fermions_vs_bosons}, it is worth noting that Eq.~\eqref{eq:collective_decay_markov} also holds for the bosonic counterpart of our model, in the case where the bath is assumed to remain approximately in its ground state, i.e.\ the vacuum state~\cite{barthel_solving_2022}.

The exact dynamics feature a number of non-Markovian effects not captured by the master equation. For instance, the master equation neglects retardation effects in the bath. In reality, we have already shown that emitted fermions propagate though the bath at a finite velocity dictated by $\Delta$ or $E_F$ (see Fig.~\ref{fig:single-impurity_bath_excitation}). Accordingly, any collective dissipative effects will be delayed by the time taken for fermions emitted at one impurity site to propagate to the next. Moreover, even after this delay, the exact dynamics will only agree with the master equation for sufficiently small impurity separations.

The collective dynamics can also show fractional decay similar to the single impurity case. This can again be captured in terms of the LTA occupation, $\overline{\mean{N_e}}$. As we show in Appendix~\ref{app:thermodynamic_LTA}, we can generalize Eq.~\eqref{eq:LTAvsGS} for initial product states of the form $\ket{\psi_0} = \ket{\psi_e}\otimes\ket{\rm FS}$, where $\ket{\psi_e}$ is an arbitrary impurity state with a well-defined number of particles, as
\begin{equation}
\label{eq:LTAvsGS_general}
\begin{split}
    \overline{\mean{O}} = \mean{O}_{\rm GS} & + \sum_{\alpha,\, \beta} \delta_{\omega_\alpha,\omega_\beta} \bra{\phi_\alpha} O \ket{\phi_\beta} \mean{\phi^\dag_\alpha \phi_\beta}_0 \\
    & - \sum_{\omega_\alpha < E_F} \bra{\phi_\alpha} O \ket{\phi_\alpha} \,.
\end{split}
\end{equation}
Here, the sums run through all single-particle bound eigenstates of $H$. In the multiple-impurity case, this includes not only BS with energies outside the band, but may also include localised dressed states emerging at the impurity energy $\Delta$, where $\Delta$ lies in the band. In recent work~\cite{leonforte_dressed_2021,leonforte_vacancy-like_2021}, a general framework has been developed for such states, which are known as \emph{bound states in the continuum} (BIC) but have also been coined \emph{vacancy-like dressed states} (VDS), since it can be shown that such states have zero amplitude on the sites to which emitters are coupled. This leads to a simple prescription for computing the wavefunction of these states, which we recall in Appendix~\ref{app:BICs}. Whether a BIC exists in the first place can be determined from the condition~\cite{leonforte_dressed_2021}
\begin{equation}
\label{eq:BIC_condition}
    \mathrm{det}\,\bm{\Sigma}(\Delta^+)=0 \,,
\end{equation}
which is a generalisation of the pole equation~\eqref{eq:pole_equation} to the multiple-impurity case, for eigenstates at energy $\Delta$.

\begin{figure}
    \centering
    \includegraphics{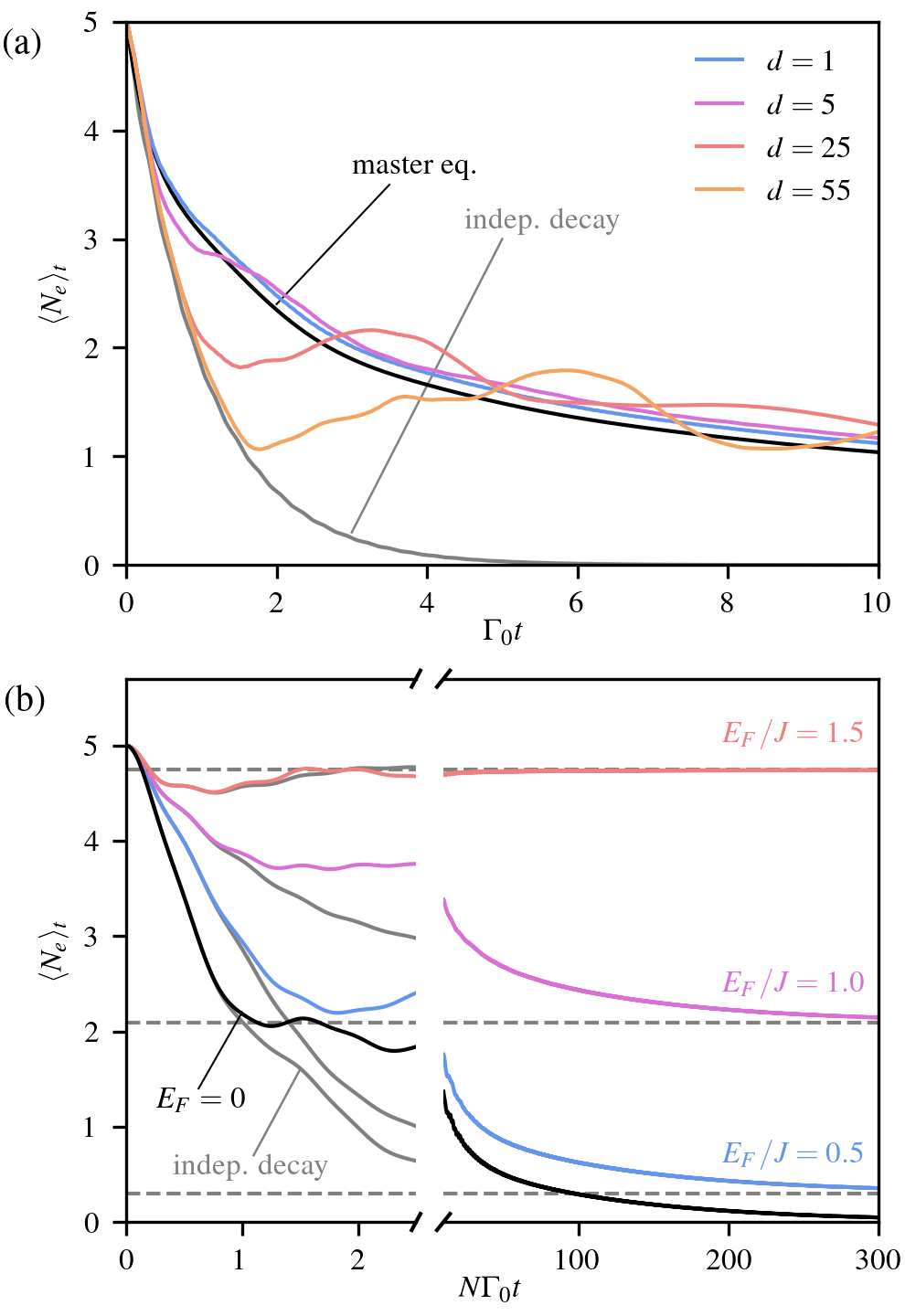}
    \caption{\emph{Collective decay.} a) Collective decay of $N=5$ equidistant impurities into an initially empty 1D bath at $\Delta=J$ and $g=0.2J$ for varied $d$ associated with the same value of $\cos(k_\Delta d)$. For reference, we also plot the master equation dynamics (black solid) and the independent decay of $N=5$ impurities (grey solid). b) Fractional decay of $N=5$ impurities at different Fermi levels, for $\Delta=J$, $g=0.4J$, and $d=5$, and independent decay of $N=5$ impurities at same values of $E_F$ (grey solid). The LTA occupation, given by Eq~\eqref{eq:LTAvsGS_general}, is also shown (grey dashed).}
    \label{fig:collective_decay}
\end{figure}

\subsection{1D bath}
As shown in Appendix~\ref{app:self-energy}, in 1D the coherent and dissipative couplings in Eq.~\eqref{eq:master_equation} read
\begin{subequations}
\label{eq:1D_mastereq_coeffs}
\begin{align}
    &J_{mn} = \frac{\Gamma_0}{2} \sin(k_\Delta\abs{r_{mn}}) \,, \\
    &\Gamma_{mn} = \Gamma_0 \cos(k_\Delta\abs{r_{mn}}) \,,
\end{align}
\end{subequations}
where $\Gamma_0$ is the single-impurity decay rate. For simplicity, we can consider configurations with the emitters equidistantly placed at a separation $d$, such that $r_{mn} = (n - m)d$. In Fig.~\ref{fig:collective_decay}a, we show the collective decay of the fully inverted state into an initially empty bath for such a configuration. While the master equation predicts the same multi-exponential decay dynamics for different values of $d$ yielding the same couplings $J_{mn}$ and $\Gamma_{mn}$, the exact dynamics display the retardation effect discussed above, and the time taken for the onset of cooperative decay intuitively increases with $d$. Note also that, as expected, the agreement between the exact and Markovian dynamics at later times is better for smaller values of $d$. For $E_F>0$, we also observe the anticipated fractional decay around the Fermi level (see Fig.~\ref{fig:collective_decay}b): $\mean{N_e}_t$ relaxes to the non-zero value~\eqref{eq:LTAvsGS_general}, computed numerically.

A particularly interesting 1D emitter configuration is the one where the emitter separation obeys $\cos(k_\Delta d) = \pm 1$. In this case, Eqs.~\eqref{eq:1D_mastereq_coeffs} imply $J_{mn} = 0$ and $\Gamma_{mn} = (\pm 1)^{\abs{m-n}}\Gamma_0$, whereby the master equation~\eqref{eq:master_equation} takes the form of the \emph{Dicke master equation}~\cite{gross_superradiance_1982}, therefore we refer to this situation as the \emph{Dicke regime}. In the case of atoms (two-level systems), the permutation-invariance of the Dicke equation together with the spin statistics leads to the characteristic scaling of the peak emission rate of the superradiant burst as $\sim N^2$~\cite{gross_superradiance_1982}. However, in our case, fermionic statistics have essentially the opposite effect: In this regime, the self-energy matrix $\bm{\Sigma}(\Delta^+)$ can be diagonalised by a unitary transformation, whereby $\mean{N_e}_t = \sum_n e^{2 \im \lambda_n t}$, with $\{\lambda_n\}_{n=1}^N$ denoting the eigenvalues of $\bm{\Sigma}(\Delta^+)$, all of which have non-positive imaginary part. In the Dicke regime, there is actually only a single non-zero eigenvalue, such that
\begin{equation}
\label{eq:Dicke_decay}
    \mean{N_e}_t=N-\left(1-e^{-N\Gamma_0t}\right)\,,
\end{equation}
implying $\lim_{t\to\infty}\mean{N_e}_t=N-1$. In other words, the dynamics only involve the emission of a single particle because the (in this case unique) bright mode can be initially occupied by at most one fermion. Again, it is worth noting that Eq.~\eqref{eq:Dicke_decay}, like the more general Eq.~\eqref{eq:collective_decay_markov}, still holds in the case where the $c_n$ are bosonic operators~\cite{navarrete-benlloch_simulating_2011,agarwal_master-equation_1970}. However, in this case the initial state is no longer Gaussian and therefore Eq.~\eqref{eq:Dicke_decay} does not capture the full evolution of the bosonic state, which in principle includes also non-trivial higher-order correlations.

\begin{figure}
    \centering
    \includegraphics[width=\linewidth]{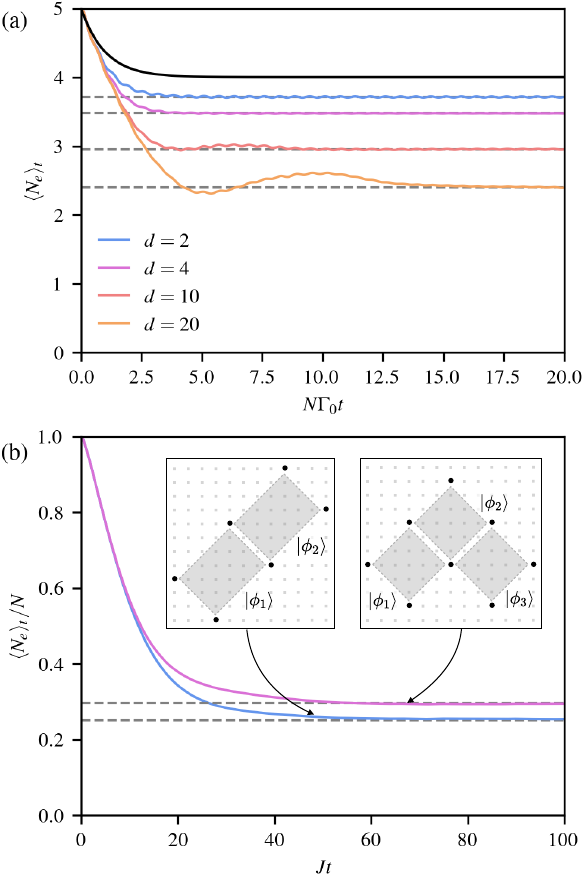}
    \caption{\emph{BIC population trapping.} a) Collective decay of $N=5$ impurities in the Dicke regime at $\Delta=J$, $E_F=0$, and $g=0.2J$ for various $d$. For reference, we also plot the master equation dynamics (black solid) and the independent decay $\mean{N_e}=Ne^{-\Gamma_0t}$ of $N$ impurities (grey solid), as well as the LTA value~\eqref{eq:Dicke_LTA} (grey dashed). b) Collective decay at $\Delta=4J$ and $g=0.2J$ for the impurity configurations shown in the insets, where  black dots indicate a site to which an impurity is coupled and grey shaded areas indicate the localised wavefunctions of the BICs, labelled by $\ket{\phi_n}$. We also plot the LTA occupations calculated in Appendix~\ref{app:BICs} (grey dashed). \vspace{-1.2em}}
    \label{fig:BIC_decay}
\end{figure}

The Markovian prediction $\overline{\mean{N_e}}=N-1$ from Eq.~\eqref{eq:Dicke_decay} can be re-framed in terms of population trapping in BICs. In 1D, Eq.~\eqref{eq:BIC_condition} takes the form
\begin{equation}
    \mathrm{det}\bm{\Sigma}(\Delta)=\prod_{n=1}^{N-1}\left(1-e^{2ik_\Delta \abs{r_n-r_{n+1}}}\right)=0\,,
\end{equation}
which implies that a BIC exists whenever two neighbouring emitters are separated by a distance $d \in (\pi/k_\Delta)\mathbb{Z}$. 
In the vacancy picture, such a pair of emitters ``cuts out'' a finite chain of $d-1$ sites, which then supports a eigenstate at energy $\Delta$ localised between the two impurity sites (see Appendix~\ref{app:BICs}). 
This implies in particular that $N-1$ such BICs emerge in the Dicke regime. 
For an empty bath ($E_F=0$), neglecting the effects of bound states outside the continuum, Eq.~\eqref{eq:LTAvsGS_general} then predicts a non-zero LTA occupation due to trapping in these BICs given by (see Appendix~\ref{app:BICs})
\begin{equation}
\label{eq:Dicke_LTA}
    \overline{\mean{N_e}}=R^2\sum_{n=1}^{N-1}\left[\frac{1+\cos\left(\dfrac{n\pi}{N}\right)}{1+R\cos\left(\dfrac{n\pi}{N}\right)}\right]^2\,,
\end{equation}
where we have defined
\begin{equation}
\label{eq:Dicke_residue}
    R=\left(1+\frac{\Gamma_0d}{2v_\Delta}\right)^{-1}\,.
\end{equation}
A formally exact expression for the case $E_F>0$ can also be obtained, but is too lengthy to reproduce here. It is worth noting, however, that for $N=2$, this more general version of Eq.~\eqref{eq:Dicke_LTA} agrees with the value of $\overline{\mean{N_e}}$ that can be obtained using an extension of the resolvent formalism for the two-impurity case (see Appendix~\ref{app:BICs}). 

We show the agreement of Eq.~\eqref{eq:Dicke_LTA} with the dynamics in Fig.~\ref{fig:BIC_decay}a. The deviation of $R$ from unity quantifies the retardation in the bath by the ratio between the timescale of the decay ($\sim1/\Gamma_0$) and the time taken for excitations to propagate between neighbouring emitters ($\sim d/2v_\Delta$)~\cite{gonzalez-tudela_markovian_2017}. For larger $d$, the population trapped in the BICs therefore decreases (see Fig.~\ref{fig:BIC_decay}). Conversely, if we neglect the retardation effects and assume $R\approx1$, Eq.~\eqref{eq:Dicke_LTA} reproduces the Markovian result $\overline{\mean{N_e}}=N-1$.

\subsection{2D bath}
In the 2D case, it is possible to find impurity configurations which ``cut out'' a full sub-lattice surrounded by vacancy-like sites in the same manner as in the 1D case. However, in 2D we also find a second, qualitatively distinct class of BIC, the first example of which was reported in Refs.~\cite{gonzalez-tudela_quantum_2017,gonzalez-tudela_markovian_2017}. 
The first distinguishing feature of these states is that they emerge only when the impurity energy is tuned to the Van Hove singularity, i.e.\ $\Delta = 4J$, where some matrix elements of $\bm{\Sigma}(\Delta^+)$ diverge. 
The second one is the fact that the corresponding impurity arrangement does not enclose a full section of the lattice. 
Instead, impurities must be placed at the vertices of a rectangle, whose sides are rotated $45^\circ$ with respect to the lattice axes. This kind of BIC has non-zero amplitude only in the interior of the rectangle (see Appendix~\ref{app:BICs}). We demonstrate the population trapping in two such configurations in Fig.~\ref{fig:BIC_decay}.

\section{Experimental considerations}
\label{sec:experiment}

\subsection{Cold-atom impurity models}
Our proposed matter-wave waveguide QED setup has been discussed in several previous works~\cite{de_vega_matter-wave_2008,navarrete-benlloch_simulating_2011,gonzalez-tudela_non-markovian_2018,stewart_analysis_2017,krinner_spontaneous_2018,stewart_dynamics_2020, lanuza_multiband_2022} therefore we will not review it in detail here. We simply recall that in these works, it was shown that by driving transitions between two atomic states which are strongly localised in isolated deep traps and delocalised across a more shallow lattice, respectively, a Hamiltonian of the form~\eqref{eq:Hamiltonian} can be realised.

Alkaline-earth-metal atoms (AEMAs) like ytterbium and strontium have emerged as arguably the most suitable for state-dependent optical lattice experiments of the kind that we envision~\cite{daley_quantum_2008}. In fact, the setup required to model our Hamiltonian, with one state localised and one delocalised, has been identified as a formidable platform for simulating complex condensed matter models~\cite{gorshkov_two-orbital_2010,foss-feig_probing_2010,foss-feig_heavy_2010} and has already been succesfully implemented in a number of experiments, both using bosonic~\cite{riegger_localized_2018} and fermionic~\cite{darkwah_oppong_observation_2019,darkwah_oppong_probing_2022} atoms.

Contrary to our model, it is important to note that such AEMA setups typically feature a (non-perturbative) on-site interaction between atoms in different internal states~\cite{gorshkov_two-orbital_2010,zhang_spectroscopic_2014}, which breaks the assumption of non-interacting particles which underlies all of our results. However, these interactions can be tuned~\cite{duan_controlling_2003,garcia-ripoll_implementation_2004}, for instance through Feschbach resonances~\cite{kohler_production_2006,bloch_many-body_2008} or by slightly displacing the traps for the two internal states from each other~\cite{bloch_many-body_2008}. Accordingly, we are confident that our proposed free-fermion matter-wave waveguide QED system is realisable in state-of-the-art experiments. We now want to establish in detail how we believe the main results that we have reported above can be observed in experiment.

\begin{figure}
    \centering
    \includegraphics{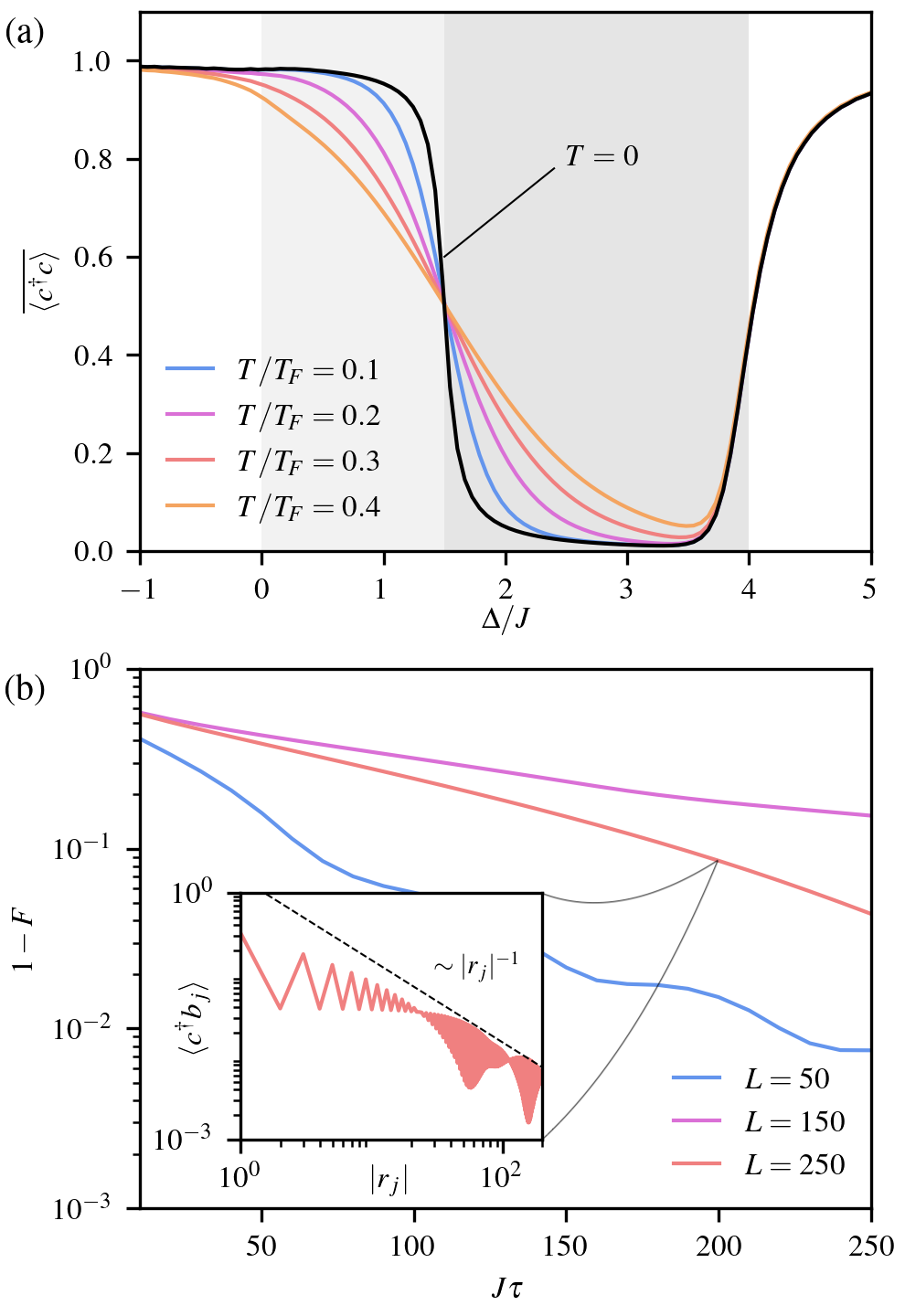}
    \caption{\textsl{Experimental considerations.} a) LTA impurity occupation for different initial bath states at temperature $T$, at $E_F=1.5J$ and $g=0.4J$, as a function of $\Delta$. For reference, we also reproduce the $T=0$ case from Fig.~\ref{fig:single-impurity_decay}c. As in Fig.~\ref{fig:single-impurity_decay}c, grey shaded areas indicate the regions $0<\Delta<E_F$ (light) and $E_F<\Delta<4J$ (dark). b) Adiabatic ground-state preparation for a single-impurity 1D system with $\Delta=E_F=2J$ and different system sizes, with the coupling linearly turned on from $g=0$ to $g=0.4J$ over a time $\tau$. For these parameters, $\xi_\mathrm{1D}=20$ in the target ground state. We plot the infidelity $1-F$, where $F$ denotes the overlap between the adiabatically varied state and the target ground state, as a function of $\tau$. We also plot the impurity-bath correlations for a finite system ($L=200$) after 200 tunneling times in the protocol, showing the cross-over from logarithmic to algebraic decay (inset).}
    \label{fig:experiment}
\end{figure}

\subsection{Quench dynamics}
Population trapping in bound states is arguably the simplest phenomenon to observe in cold-atom experiments, as it only requires initialising a few atoms in the deeply trapped internal state at selected positions in the optical lattice.
Indeed, at the single-excitation level, experimental confirmation of the effect of a bound state on the single-impurity matter-wave dynamics has already been reported~\cite{stewart_dynamics_2020}.
It should therefore be possible to observe the population trapping of multiple excitations in the bound states (both outside and in the continuum) of multiple-impurity systems, as shown in Fig.~\ref{fig:BIC_decay}. 
It should also be possible to observe the collective decay effects (multi-exponential decay) and non-Markovian dynamics (retardation) shown in Fig.~\ref{fig:collective_decay} as the distance between the impurities is increased.

To explore fractional decay around the Fermi level for one or multiple impurities, we need to prepare the more complex initial state $\ket{\psi_0} = \left(\prod_n c_n^\dag\right) \ket{\rm FS}$. Of course, the true zero-temperature Fermi sea state cannot be realised in experiment. Instead, the state $\ket{\psi_0}$ would be characterised by the Fermi-Dirac distribution, i.e.\ $\mean{b_\vec{k}^\dag b_\vec{q}}_0=\delta_{\vec{k}\vec{q}}f(\omega_\vec{k})$ with $f(\omega)=\left(1+e^{(\omega-E_F)/k_BT}\right)^{-1}$ at a temperature $T$. Such an initial state would no longer be a product state and therefore most of the discussion regarding dynamics in Sec.~\ref{sec:formalism} would no longer apply. However, we can estimate the effect of this different initial state on our results by numerical evaluation of Eq.~\eqref{eq:correlation_matrix}. For instance, Fig.~\ref{fig:experiment}a shows the single-impurity LTA occupation, complementary to Fig.~\ref{fig:single-impurity_decay}c, for such a finite-temperature initial state. Clearly, for sufficiently low temperatures, the LTA occupation still qualitatively resembles the zero-temperature profile. In recent experiments, temperatures $T/T_F\sim0.2$ (with $T_F=E_F/k_B$) have already been reported~\cite{sonderhouse_thermodynamics_2020, milner_high-fidelity_2023} and the phenomenon of fractional decay around the Fermi level should therefore certainly be observable in experiment.

\subsection{Ground-state correlations}

\subsubsection{Measurement schemes}
Another relevant question is the feasibility of measuring single-impurity ground-state correlations like the ones shown in Fig.~\ref{fig:single-impurity_correlations}. In particular, we have discussed how the real-space structure of the impurity screening cloud can be inferred from the correlator $\mean{c^\dag b_j}_\mathrm{GS}$. The experimental challenge of measuring this correlator directly can be circumvented by exploiting the Gaussianity of our model: Wick's Theorem implies
\begin{equation}
    \abs{\mean{c^\dag b_j}_\mathrm{GS}}=\sqrt{\mean{c^\dag c\,b_j^\dag b_j}_\mathrm{GS}-\mean{c^\dag c}_\mathrm{GS}\mean{b_j^\dag b_j}_\mathrm{GS}}\,,
\end{equation}
and both terms on the right hand side of this expression can be obtained through site-resolved density measurements of the kind utilised in well-established cold-atom microscope experiments~\cite{cheuk_quantum-gas_2015, parsons_site-resolved_2015, sherson_single-atom-resolved_2010, haller_single-atom_2015}. While we find that the phase of the correlations $\mean{c^\dag b_j}$ cannot be reconstructed even from higher-order correlations, the magnitude $\abs{\mean{c^\dag b_j}}$ is sufficient to characterise the screening.

On the other hand, the full ground-state correlations, including also their complex phase, could be measured using an alternative approach similar to the photoemission spectroscopy scheme proposed in Ref.~\cite{bohrdt_angle-resolved_2018}: 
by introducing local modulations of the bath lattice at the impurity site and another site $j$, a small number of fermions (less than one) can be resonantly excited into an auxiliary, detection lattice, which is initially empty, and whose chemical potential is offset from the original lattice by an energy much larger than the inter-lattice tunelling. 
The collective momentum-space modes of this auxiliary lattice right after the modulation (at $t = 0$) can then be expressed in terms of the original bath operators as $\tilde{b}_\vec{k} = b_0 + e^{i\vec{k}\cdot\vec{r}_j}b_j$. 
After a sufficiently long period of ballistic expansion in the detection lattice, the occupation of these modes can be measured using a standard time-of-flight protocol~\cite{greiner_exploring_2001}. 
Writing $\mean{b_0^\dagger b_j}_{t=0} = \mean{b_0^\dagger b_j}_\mathrm{GS} = \abs{\mean{b_0^\dagger b_j}_\mathrm{GS}}e^{i\varphi_j}$, the measured occupations then read
\begin{multline}
    \mean{\tilde{b}_\vec{k}^\dagger\tilde{b}_\vec{k}}_{t = 0} = \mean{b_0^\dagger b_0}_\mathrm{GS}+\mean{b_j^\dagger b_j}_\mathrm{GS} \\ +2\abs{\mean{b_0^\dagger b_j}_\mathrm{GS}}\cos\left(\vec{k}\cdot\vec{r}_j+\varphi_j\right)\,,
\end{multline}
allowing us to infer both $\abs{\mean{b_0^\dagger b_j}_\mathrm{GS}}$ and $\varphi_j$ from the obtained data for (many) different values of $\vec{k}$. 
While the above discussion focuses on bath-bath correlations, it stands to reason that a similar method could also be used to measure the phase of $\mean{c^\dagger b_j}_\mathrm{GS}$, by coupling only the impurity site (state) and not the bath site to which it is coupled to the auxiliary detection lattice.

\subsubsection{Ground-state preparation}
The more significant experimental challenge is the preparation of the many-body ground state itself. A well-established approach in this context is to start from a different, simpler many-body state and adiabatically tune the parameters of the system to transform this state into the desired ground state~\cite{albash_adiabatic_2018,garcia-ripoll_implementation_2004,trebst_d-wave_2006,sorensen_adiabatic_2010}. In our case, this would involve preparing the Fermi sea state and adiabatically turning on the coupling $g$. 
Crucially, the preparation time depends on the size of the energy gap between the ground state and the first excited state during the protocol~\cite{albash_adiabatic_2018}, which for our model is given by $\sim\pi v_F/L^d$. For large systems, it therefore becomes unfeasible to adiabatically prepare the ground state. 
This is exacerbated by the fact that the overlap between any two ground states of impurity models with different coupling strengths $g$ decays with the system size, a phenomenon known as the orthogonality catastrophe~\cite{anderson_infrared_1967,gebert_andersons_2014}. However, the cross-over from logarithmic to algebraic scaling of the correlations can already be observed in comparatively small finite systems, which can be adiabatically prepared in reasonable time compared to experimental timescales (see Fig.~\ref{fig:experiment}b).

An alternative way to observe the impurity screening cloud is given by Eqs.~\eqref{eq:LTAvsGS} and~\eqref{eq:LTAvsGS_general}: In a finite system with $\Delta$ sufficiently deeply in the band, the initial state $\ket{\psi_0}=c^\dag\ket{\rm FS}$ relaxes to a state sufficiently resembling the ground state with the same number of excitations which displays the desired impurity screening effect.

\section{Conclusion}
In this work, we have studied the physics of a spinless fermionic system consisting of one or more impurities coupled to a single-band bath in 1D and 2D. By extending some of the tools originally developed in quantum optics to address the properties of quantum emitters coupled to structured baths, we have been able to incorporate many effects that have typically been neglected in condensed-matter studies of such systems~\textendash~for example, the effect of band edges and impurity-bath bound states. Conversely, from the perspective of quantum optics, our work provides an insight into the effect of many-body physics and fermionic particle statistics on typical quantum-optical phenomena, such as particle emission, fractional decay, effective bath-mediated interactions, and collective effects such as sub- and superradiance.

Specifically, for a single impurity, we presented exact expressions for the two-point correlation functions which fully capture the ground state. 
This allowed us to characterise in detail the real-space signatures of an impurity screening cloud in our model. 
In 1D, we showed that the impurity-bath correlations obey a universal scaling law over a much larger parameter range than previously expected.
More generally, in arbitrary dimension $d$, we demonstrated that the impurity-bath correlations have the same long-range algebraic scaling as the bath-bath correlations in a bath without an impurity, as $\sim \abs{\vec{r}}^{-(d+1)/2}$ for $\abs{\vec{r}}\to\infty$.

Furthermore, we studied the quench dynamics of an initial state with the impurity populated and a Fermi sea state in the bath. We unveiled qualitatively different emission of matter waves from the impurity into the bath depending on the value of the impurity energy relative to the Fermi level, and we derived a formula capturing the steady state after emission, which clarifies the effect of single-particle bound states on the dynamics and characterises when the system relaxes to its ground state. For multiple impurities, we showed that the quench dynamics of an analogous initial state, in which all impurities are occupied and the bath is in a Fermi sea state, typically display multi-exponential decay of excitations into the bath. We also characterized the steady state of the system and once again highlighted the role of single-particle bound states, which, for certain impurity configurations, can also emerge at energies in the continuum. In 1D, we related the fractional decay in the Dicke regime (i.e. the regime with only a single collective ``bright'' mode) to the existence of such BICs, which trap the excitations in the impurity modes. In 2D, we also discussed population trapping in BICs, generalizing examples of BICs previously reported in the literature.

Importantly, some of these effects, such as the signatures of single-impurity screening cloud, fractional decay due to Fermi statistics, and population trapping in BICs, could be observed with state-of-the-art cold atom experiments, which allow for the exploration of regimes that are typically inaccessible in more conventional solid-state setups.

Moreover, our proposed scenario of fermionic matter-wave quantum optics opens the door to a number of interesting avenues of further research. While we have focused so far on spinless non-interacting models, we expect that including spin and interactions will reveal even more intriguing phenomena intrinsically linked to the fermionic character of the excitations. In particular, the scenario of collective dissipative dynamics of interacting fermionic impurities constitutes a potentially interesting intersection between many-body physics and quantum optics. In the superlative, it could prove fruitful to study the quantum optics of Kondo-type impurities coupled to structured reservoirs on the one hand, and emitters in strongly-correlated baths on the other hand. However, moving beyond our simple models could certainly constitute a considerable computational challenge requiring a more sophisticated theoretical toolbox than we have presented here.

\section{Acknowledgements}
We are grateful to S. Blatt for helpful discussions and advice. B.W., M.B., and J.I.C. acknowledge funding from the projects FermiQP and EQUAHUMO of the Bildungsministerium f{\"u}r Bildung und Forschung (BMBF), as well as from the Munich Center for Quantum Science and Technology (MCQST), funded by the Deutsche Forschungsgemeinschaft (DFG) under Germany's Excellence Strategy (EXC2111 - 390814868). ED acknowledges support from the ARO Grant No. W911NF-16-1-0361 and the SNSF project 200021\_212899.

\section{Code availability}
The computer codes used to obtain the results presented in this work are publicly available at \href{https://doi.org/10.5281/zenodo.8096145}{https://doi.org/10.5281/zenodo.8096145}

\appendix

\section{Self-energy integrals}
\label{app:self-energy}
In this appendix, we show how the evaluation of the self-energy integral~\eqref{eq:self-energy} can be approached analytically.

\subsection{Analytical expression in 1D}
Defining $\beta=e^{ik}$, the 1D self-energy function can be re-written as the contour integral
\begin{equation}
\label{eq:self-energy_contour}
    \Sigma(z,r) = \frac{g^2}{J} \oint_{\abs{\beta}=1} \frac{d\beta}{2\pi i} \frac{\beta^{\abs{r}}}{p(\beta; z)}
\end{equation}
where the denominator is the second-order (palindromic) polynomial $p(\beta; z) = \beta^2 + (z/J - 2)\beta + 1$, and the integration contour is the unit circumference traversed in a counter-clockwise manner. 
The polynomial can be factorised as $p(\beta; z) = [\beta - \beta_+(z)][\beta - \beta_-(z)]$ with
\begin{equation}
    \beta_\pm(z) \equiv 1 - \frac{z}{2J} \pm \sqrt{\left(\frac{z}{2J} - 1\right)^2 - 1}\,.
\end{equation}
It is easy to see that the roots satisfy
\begin{subequations}
\begin{align}
    \beta_+(z) \beta_-(z) & = 1 \,, \label{eq:roots_product} \\
    - [\beta_+(z) + \beta_-(z)] & = z/J - 2 \,. \label{eq:roots_sum}
\end{align}
\end{subequations}
From \eqref{eq:roots_product}, we realize that the roots are inverses of each other. 
Consequently, only one of them contributes to the integral, which we denote by $\beta_{\rm in}$, while we denote the other one as $\beta_{\rm out}$. 
Hence,
\begin{equation}
\label{eq:self-energy1D}
    \Sigma(z, r) = \frac{g^2 \beta^{\abs{r}}_\mathrm{in}(z)}{J[\beta_\mathrm{in}(z) - \beta_\mathrm{out}(z)]} \,.
\end{equation}
Furthermore, for $0 < \Delta < 4J$, it can be shown that $\abs{\beta_{\rm in}(\Delta^+)} = 1$. 
In that regime, we can choose $\beta_{\rm in} = e^{i\phi}$ and $\beta_{\rm out} = e^{-i\phi}$.
Now, Eq.~\eqref{eq:roots_sum} implies $\phi = \arccos(1 - \Delta/(2J)) \equiv k_\Delta$. 
Substituting back into Eq.~\eqref{eq:self-energy1D},
\begin{equation}
\label{eq:1D_self-energy}
    \Sigma(\Delta^+, r) = \frac{g^2 e^{ik_\Delta\abs{r}}}{2Ji\sin(k_\Delta)} = -i \frac{\Gamma_0}{2} e^{i k_\Delta \abs{r}} \,.
\end{equation}
As can be seen from the first equality, the single-emitter self energy $\Sigma_e(\Delta^+) \equiv \Sigma(\Delta^+, 0)$ is purely imaginary for values of $\Delta$ within the band range, so in the second equality we express the general self-energy function in terms of the single-impurity decay rate $\Gamma_0$. 
This formula allows us to evaluate the coherent couplings and collective decay rates appearing in the master equation~\eqref{eq:master_equation} in a straightforward manner, leading to Eqs.~\eqref{eq:1D_mastereq_coeffs}.

\subsection{Relation to 1D integral in 2D}
In the 2D case, where $\vec{r} = (r^x, r^y)$, the self-energy integral can be simplified by integrating over one momentum-space component in the manner outlined above to obtain
\begin{equation}
    \Sigma(z, \vec{r}) = \frac{g^2}{J} \int_{-\pi}^\pi\frac{dk}{2\pi} \frac{e^{ikr^y}\,\beta^{\abs{r^x}}_\mathrm{in}(k, z)}{\beta_\mathrm{in}(k, z) - \beta_\mathrm{out}(k, z)} \,,
\end{equation}
where $\beta_\mathrm{in}(k, z)$ and $\beta_\mathrm{out}(k, z)$ denote the roots of the polynomial $p(\beta; k, z) = \beta^2 + (z/J - 4 + 2\cos k)\beta + 1$, with $\abs{\beta_{\rm in}} < 1$ and $\abs{\beta_{\rm out}} > 1$. 
While the remaining integration cannot in general be performed analytically, it does make the evaluation of $\Sigma(z, \vec{r})$ less numerically demanding than performing the full 2D integral.

\section{Fermionic master equation}
\label{app:master_equation}
Here, we derive the fermionic master equation, Eq.~\eqref{eq:master_equation}, by formally tracing out the bath modes. 
The starting point is the time-local, Born-Markov master equation \cite{BreuerPetruccione2007}
\begin{equation}
    \dot{\rho}_I(t) = - \int_0^\infty dt' \tr_B \left\{[V_I(t), [V_I(t - t'), \rho_I(t)\otimes \rho_B]]\right\} \,,
\end{equation}
expressed here in the interaction picture,
\begin{equation*}
    O_I(t) = e^{i(H_S + H_B)t} O e^{-i(H_S + H_B)t} \,,
\end{equation*}
and where we take $\rho_B = \ket{\rm FS}\!\bra{\rm FS}$. This equation can be derived formally using projection operators~\cite{schuetz_superradiance-like_2012}. Qualitatively, the underlying assumption is that the bath correlation time is short compared to the characteristic timescale of the impurity-bath interaction $g$, implying that the bath remains approximately in thermal equilibrium despite its coupling to the impurity~\cite{BreuerPetruccione2007,schuetz_superradiance-like_2012}. Expanding the nested commutators we can express
\begin{equation}
    \dot{\rho}_I(t) = \int_0^\infty dt' \left[T_1(t,t') - T_2(t,t') + \mathrm{H.c.}\right] \,,
\end{equation}
where we have defined
\begin{align}
    T_1(t, t') & = \tr_B \left\{V_I(t - t') \rho_I(t)\otimes\rho_B V_I(t) \right\} \,, \\
    T_2(t, t') & = \tr_B \left\{V_I(t) V_I(t - t') \rho_I(t) \otimes \rho_B\right\} \,.
\end{align}
With the explicit form of $H_S\,,H_B\,,$ and $V$ given in Eqs.~\eqref{eq:Hamiltonian} and~\eqref{eq:Hamiltonian_momentum}, it is easy to see that
\begin{equation}
    V_I(t) = \frac{g}{\sqrt{L^d}}\sum_{n,\, \vec{k}} \left[c_n^\dagger b_\vec{k}\,e^{i\vec{k}\cdot\vec{r}_n} e^{i(\Delta - \omega_\vec{k})t} + \mathrm{H.c.}\right] \,.
\end{equation}
Substituting this into $T_1(t,t')$ and $T_2(t,t')$, using the fact that for the particular bath state considered $\tr_B(b_\vec{k} b^\dag_\vec{q} \rho_B) = \delta_{\vec{k}\vec{q}} \Theta(\omega_\vec{k} - E_F)$ and $\tr_B(b^\dag_\vec{k} b_\vec{q} \rho_B) = \delta_{\vec{k}\vec{q}} \Theta(E_F - \omega_\vec{k})$, while $\tr_B(b^\dag_\vec{k} b^\dag_\vec{q} \rho_B) = \tr_B(b_\vec{k} b_\vec{q} \rho_B) = 0$, we have
\begin{widetext}
\begin{align}
    T_1(t, t') & = g^2 \sum_{m,\, n} \left[c^\dag_n \rho_I(t) c_m \int\limits_{\omega_\vec{k} < E_F} \frac{d^d\vec{k}}{(2\pi)^d} \, e^{-i\vec{k}\cdot\vec{r}_{mn}} e^{-i(\Delta - \omega_\vec{k})t'} + c_n \rho_I(t) c^\dag_m \int\limits_{\omega_\vec{k} > E_F} \frac{d^d\vec{k}}{(2\pi)^d} \, e^{i\vec{k}\cdot\vec{r}_{mn}} e^{i(\Delta - \omega_k)t'}\right] \,, \\
    T_2(t, t') & = g^2 \sum_{m,\, n} \left[c^\dag_m c_n \rho_I(t) \int\limits_{\omega_\vec{k} < E_F} \frac{d^d\vec{k}}{(2\pi)^d}\, e^{-i\vec{k}\cdot\vec{r}_{mn}} e^{-i(\Delta - \omega_\vec{k})t'} + c_m c^\dag_n \rho_I(t) \int\limits_{\omega_\vec{k} > E_F} \frac{d^d\vec{k}}{(2\pi)^d} \, e^{i\vec{k}\cdot\vec{r}_{mn}} e^{i(\Delta - \omega_\vec{k})t'}\right] \,.
\end{align}
\end{widetext}
The integrals in time can be performed substituting $\Delta \to \Delta \pm i \epsilon$, and then taking the limit $\epsilon \to 0$,
\begin{equation}
\label{eq:selfe_gtr}
    \begin{split}
        & g^2 \int_0^\infty dt' \int\limits_{\omega_\vec{k} > E_F} \frac{d^d\vec{k}}{(2\pi)^d} \, e^{i\vec{k}\cdot\vec{r}_{mn}} e^{i(\Delta^+ - \omega_\vec{k})t'} \\
        & \qquad = i g^2 \int\limits_{\omega_\vec{k} > E_F} \frac{d^d\vec{k}}{(2\pi)^d} \, \frac{e^{i\vec{k}\cdot\vec{r}_{mn}}}{\Delta^+ - \omega_\vec{k}} = i\left(J^>_{mn} - i\frac{\Gamma^>_{mn}}{2}\right) \,,
    \end{split}
\end{equation}
and similarly,
\begin{equation}
\label{eq:selfe_less}
    \begin{split}
        & g^2 \int_0^\infty dt' \int\limits_{\omega_\vec{k} < E_F} \frac{d^d\vec{k}}{(2\pi)^d} \, e^{-i\vec{k}\cdot\vec{r}_{mn}} e^{-i(\Delta^- - \omega_\vec{k})t'} \\
        & \qquad = -i g^2 \int\limits_{\omega_\vec{k} < E_F} \frac{d^d\vec{k}}{(2\pi)^d} \, \frac{e^{i\vec{k}\cdot\vec{r}_{mn}}}{\Delta^- - \omega_\vec{k}} = -i\left(J^<_{mn} + i\frac{\Gamma^<_{mn}}{2}\right) \,.
    \end{split}
\end{equation}
In the last step, we have used the Sokhotski-Plemelj theorem, expressing the end result as the sum of two contributions: one coming from the principal value of the integral, $J^\lessgtr_{mn}$, and another one stemming from the Dirac delta distribution, $\Gamma^\lessgtr_{mn}/2$.
The fact that the dispersion relation is symmetric under the change $\vec{k}\to -\vec{k}$ guarantees that both $J^\lessgtr_{mn}$ and $\Gamma^\lessgtr_{mn}$ are real and symmetric, $J^\lessgtr_{mn} = J^\lessgtr_{nm}$ and $\Gamma^\lessgtr_{mn} = \Gamma^\lessgtr_{nm}$.
Also, comparing Eqs.~\eqref{eq:selfe_gtr} and \eqref{eq:selfe_less} with the definition of the self-energy function, Eq.~\eqref{eq:self-energy}, it becomes clear that Eq.~\eqref{eq:gamma_lessgtr} holds. Putting everything together,
\begin{equation}
\begin{split}
    & \dot{\rho}_I = \sum_{m,\, n}\Biggl[\left(-i J^<_{mn} + \frac{\Gamma^<_{mn}}{2}\right)\left(c^\dag_n \rho_I c_m - c^\dag_m c_n \rho_I\right) \\
    & \quad\qquad + \left(i J^>_{mn} + \frac{\Gamma^>_{mn}}{2}\right)\left(c_n \rho_I c^\dag_m - c_m c^\dag_n \rho_I\right) + \mathrm{H.c.}\Biggr] \,.
\end{split}
\end{equation}
Last, it is a matter of algebra to show that
\begin{equation}
\begin{split}
    \dot{\rho}_I & = -i \sum_{m,\, n} (J^<_{mn} + J^>_{mn})[c^\dag_m c_n, \rho_I] \\
    & \qquad\qquad + \mathcal{D}_> \rho_I + \mathcal{D}_< \rho_I \,.
\end{split}
\end{equation}
Going back to the Schrödinger picture, noting also that $J^<_{mn} + J^>_{mn} = J_{mn}$, this then yields the result presented in the main text.

\section{Ground-state correlations}
\label{app:screening}

\subsection{1D bath}
In the case $d = 1$, Eq.~\eqref{eq:GS_cloud} can be simplified using the expression for the self-energy function introduced in Appendix~\ref{app:self-energy}. 
It then reads, neglecting the bound state contribution in anticipation of the wide-band approximation,
\begin{equation}
    \mean{c^\dag b_j}_{\rm GS} = \frac{g}{\pi} \re\left[\int_0^{k_F} dk\, G_e(\omega_k^+) e^{ik\abs{r_j}}\right] \,.
\end{equation}
Under a wide-band approximation, we linearise the dispersion relation around the Fermi level as $\omega_k \approx E_F + v_F(\abs{k} - k_F)$, where $v_F$ denotes the group velocity at the Fermi level. 
We also assume $\Sigma_e(\omega_k^+) \approx \Sigma_e(E_F^+) = -ig^2/v_F$, so that, substituting $z = k - k_F$,
\begin{equation}
    \mean{c^\dag b_j}_{\rm GS} = \frac{g}{\pi v_F} \re\left[e^{ik_F\abs{r_j}}\int_{-k_F}^0 dz\, \frac{e^{iz\abs{r_j}}}{z - z_0}\right] \,,
\end{equation}
where $z_0 = (\Delta - E_F)/v_F - i(g/v_F)^2$. 
For large $k_F$, the integrand is highly oscillatory and hence we can extend the lower limit of integration to $-\infty$. 
This allows us to evaluate the integral analytically as
\begin{equation}
    \mean{c^\dag b_j}_{\rm GS} = \frac{g}{\pi v_F} \re\left[e^{ik_F\abs{r_j}} f(z_0 \abs{r_j})\right] \,,
\end{equation}
where
\begin{equation}
    f(z) \equiv e^{iz} \left[\Ci(z) + i\left(\frac{\pi}{2} - \Si(z)\right)\right] \,;
\end{equation}
$\Ci(z)$ and $\Si(z)$ are the cosine and sine integrals, respectively~\cite{AbramowitzStegun1964}. 
In addition, we find that $f(z_0 \abs{r_j})$ is approximately independent of the argument of $z_0$, 
so that we can replace $z_0 \abs{r_j}$ by $\abs{r_j}/\xi_\mathrm{1D}$ with $\xi_\mathrm{1D} = \abs{z_0}^{-1}$, as defined in Eq.~\eqref{eq:1D_screening_length}. Noting that $\re[e^{ik_F\abs{r_j}} f(\abs{r_j}/\xi_{\rm 1D})] \leq \abs{f(\abs{r_j}/\xi_{\rm 1D})}$, we can simply define the envelope of the correlations as $F(\abs{r_j}) = \abs{f(\abs{r_j})}$.

\subsection{2D bath}

\subsubsection{Efficient numerical computation}
Neglecting the LBS contribution, Eq.~\eqref{eq:GS_cloud} can be cast in the form of a one-dimensional Fourier transform.
To do so, one has to perform a change of variables, expressing the integral in terms of the momentum parallel and perpendicular to $\vec{r}_j$.
More concretely, in 2D, for $\vec{r}_j = (m, n) r$ we introduce $\vec{k}' \equiv (k, k_\perp) = T(k_x, k_y)$, where $T$ is a conformal linear transformation,
\begin{equation}
    \begin{pmatrix} k \\ k_\perp \end{pmatrix} = \frac{1}{m^2 + n^2} \begin{pmatrix} m & n \\ -n & m \end{pmatrix} \begin{pmatrix} k_x \\ k_y \end{pmatrix}\,.
\end{equation}
Thus, the first term in Eq.~\eqref{eq:GS_cloud} becomes
\begin{multline}
    \int_{\rm BZ} \frac{d^2 \vec{k}}{(2\pi)^2} \, G_e(\omega_\vec{k}^-) e^{i\vec{k}\cdot\vec{r}_j} \Theta(E_F - \omega_\vec{k}) \\ = \int_0^{2\pi} \frac{dk}{2\pi} \, e^{ik r'} f_1(k) \,,
\end{multline}
where we have defined the one-dimensional integral
\begin{equation}
\label{eq:def_f1}
    f_1(k) = \int_0^{2\pi} \frac{dk_\perp}{2\pi} \, G_e(\Omega_{\vec{k}'}^-) \Theta(E_F - \Omega_{\vec{k}'})\,.
\end{equation}
Here, $r' \equiv (m^2 + n^2) r$ and $\Omega_{\vec{k}'} \equiv \omega_{T^{-1}(\vec{k})}$. Note that the original integration region, $\mathrm{BZ}$, is the square with vertices $2\pi\times\{(0, 0), (1, 0), (0, 1), (1, 1)\}$, whereas $T(\mathrm{BZ})$ is a scaled and rotated version of it.
Since the integrand is periodic (invariant under translations by reciprocal lattice vectors) we could have considered instead a different integration region, $\mathrm{BZ}'$, the square with vertices $2\pi\times\{(0, 0), (m, n), (-n, m), (m - n, m + n)\}$, which has an area $(m^2 + n^2)$ times that of the original $\mathrm{BZ}$. 
The transformed integration region $T(\mathrm{BZ}')$ coincides with $\mathrm{BZ}$. 

Similarly, we can do the same change of variables to express the self-energy function as
\begin{equation}
    \Sigma(z, \vec{r}_j) = g^2 \int_0^{2\pi} \frac{dk}{2\pi} \, e^{ik r'} \int_0^{2\pi} \frac{dk_\perp}{2\pi} \, \frac{1}{z - \Omega_{\vec{k}'}}\,.
\end{equation}
Assuming $m \geq n > 0$, calling $x = e^{ik}$ and $y = e^{ik_\perp}$, we can integrate out $k_\perp$ using the residue integration technique, obtaining
\begin{equation}
    \Sigma(z, \vec{r}_j) = \int_0^{2\pi} \frac{dk}{2\pi} \, e^{ik r'} \sigma(z, k) \,,
\end{equation}
with $\sigma(z, k)$ defined as
\begin{equation}
    \sigma(z, k) = \frac{g^2}{J} \sum_{\substack{p(y_j) = 0,\\ \text{s.t.}\, \lvert y_j\rvert < 1}} y^{m - 1}_j \mathrm{Res}\left(\frac{1}{p(y;z,x)}, y_j\right) \,,
\end{equation}
where $p(y;z,x)$ is the $2m$-order polynomial $p(y;z,x)=x^n y^{2m} + x^{-m}y^{m + n} + (z/J - 4)y^m + x^m y^{m - n} + x^{-n}$ with roots $y=y_j$. 
Note that if $y_j$ is a simple root, the residue is simply given by $\mathrm{Res}\left(1/p(y;z,x), y_j\right) = 1/p'(y_j;z,x)$. Thus, the second term in Eq.~\eqref{eq:GS_cloud} can be written as
\begin{multline}
    \int_{\rm BZ} \frac{d^2 \vec{k}}{(2\pi)^2} \, \abs{G_e(\omega_\vec{k}^+)}^2\Sigma(\omega_\vec{k}^+, \vec{r}_j) \Theta(E_F - \omega_\vec{k}) \\ = \int_0^{2\pi} \frac{dk}{2\pi} \, e^{ik r'} f_2(k) \,,
\end{multline}
with a new auxiliary one-dimensional integral defined as
\begin{align}
\label{eq:def_f2}
    f_2(k) & = g^2 \int_0^{2\pi} \frac{dk_\perp}{2\pi} \int_{\rm BZ} \frac{d^2\vec{q}}{(2\pi)^2} \, \frac{\abs{G_e(\omega_\vec{q}^+)}^2}{\omega_\vec{q}^+ - \Omega_{\vec{k}'}} \Theta(E_F - \omega_\vec{q}) \\
    & = \int_0^{E_F} d\epsilon\, \rho(\epsilon)\abs{G_e(\epsilon^+)}^2 \sigma(\epsilon^+, k)\,.
\end{align}
Finally, we arrive at the compact expression
\begin{equation}
    \label{eq:correlations_ft}
    \mean{c^\dag b_j}_{\rm GS} \approx g \int_0^{2\pi} \frac{dk}{2\pi} e^{i k r'} \re\left[f_1(k) + f_2(k)\right] \,,
\end{equation}
which follows from the fact that the correlations are real, since the total Hamiltonian is real, and $f_n(-k) = f_n(k)$ for $n=1,2$.

For the special case of $(m, n) = (1, 0)$, we can find an analytical expression for $\sigma(z, k)$, which reads
\begin{equation}
    \sigma(z, k) = \frac{g^2}{J} \frac{\sign(\re(b))}{\sqrt{b^2 - 4}} \,, \ b = \frac{z}{J} - 2\cos(k) - 4 \,,
\end{equation}
while for $(m, n) = (1, 1)$, we find that
\begin{equation}
    \sigma(z, k) = \frac{g^2}{2J\cos(k)} \frac{\sign(\re(b))}{\sqrt{b^2 - 4}} \,, \ b = \frac{z - 4J}{2J\cos(k)} \,.
\end{equation}
In practice, for general $\vec{r}=(m,n)$ we compute $\re f_n(k)$ numerically, and then evaluate Eq.~\eqref{eq:correlations_ft} using a Fast Fourier Transform algorithm.

Note that the same approach can be used to compute the bath-bath correlations~\eqref{eq:GS_bath_correlations} in an efficient way, and this is how we obtain Fig.~\ref{fig:single-impurity_correlations}c.

\subsubsection{Long-range scaling}
The long-range behavior of the Fourier transform is determined by the singularities of the function to be transformed.
The singularities of $\re f_1(k)$ arise because the integrand in its definition, Eq.~\eqref{eq:def_f1}, is discontinuous at the Fermi surface, $\Omega_{\vec{k}'} = E_F$. 
Similarly, the integrand in the definition of $\re f_2(k)$ (after performing the integral in the $q$-variables) has a logarithmic divergence at the Fermi surface.
Thus, if the Fermi surface is smooth, after integrating along the perpendicular momentum $k_\perp$, $\re [f_1(k) + f_2(k)]$ has a set of singularities of the form $\re [f_1(k) + f_2(k)] \simeq A_i + B_i \abs{k - k_{F,i}}^{1/2}$ for $k \to k_{F,i}$, which results in a long-range scaling of the correlations as $\sim r^{-3/2}$ \cite{Bender1999}.
In fact, the singularities are of the same type as the ones one would obtain following a similar procedure for the bath-bath correlations in an impurity-free bath, which can be computed as \cite{Wen2007}
\begin{equation}
    \mean{b^\dag_0 b_j}_{\rm FS} = \int_{\rm BZ} \frac{d^2 \vec{k}}{(2\pi)^2} \, e^{i\vec{k}\cdot\vec{r}_j} \Theta(E_F - \omega_\vec{k}) \,.
\end{equation}
A similar analysis can be performed for higher dimensions, arriving at the same conclusion.

\section{Thermodynamic LTA}
\label{app:thermodynamic_LTA}

A general eigenstate of the Hamiltonian~\eqref{eq:Hamiltonian} with energy $E$ can be written as
\begin{equation}
    \ket{\phi_\alpha(E)}=\left(\sum_n a_{\alpha n}c_n^\dagger+\sum_\vec{k}a_{\alpha\vec{k}}b_\vec{k}^\dagger\right)\ket{\rm vac}\,,
\end{equation}
where $\alpha = 1, \ldots, D(E)$ labels the degenerate eigenstates at energy $E$ and $D(E)$ denotes the associated degeneracy. 
The associated eigenvalue equation $H \ket{\phi_\alpha(E)} = E \ket{\phi_\alpha(E)}$ implies that
\begin{subequations}
\begin{align}
    \Delta \vec{a}_\alpha + \sum_\vec{k} \vec{g}_\vec{k} a_{\alpha\vec{k}} & = E \vec{a}_\alpha \,, \\
    \vec{g}^\dag_\vec{k} \vec{a}_\alpha + \omega_\vec{k} a_{\alpha\vec{k}} & = E a_{\alpha\vec{k}}\,,
\end{align}
\end{subequations}
where we have defined $\vec{a}_\alpha = (a_{\alpha 1}, \ldots, a_{\alpha N})^T$ and $\vec{g}_\vec{k} = g L^{d/2}(e^{-i\vec{k}\cdot\vec{r}_1}, \ldots, e^{-i\vec{k}\cdot\vec{r}_N})^T$. 
By the second equation,
\begin{equation}
\label{eq:impurity_amplitudes}
    a_{\alpha\vec{k}} = \frac{\bm{g}_\vec{k}^\dagger\vec{a}_\alpha}{E-\omega_\vec{k}}
\end{equation}
and substituting this solution into the first equation,
\begin{equation}
    \left[E-\Delta-\bm{\Sigma}(E)\right]\vec{a}_{\alpha}=0\,,
\end{equation}
where we have introduced
\begin{equation}
\label{eq:self-energy_matrix}
    \bm{\Sigma}(z)\equiv\sum_\vec{k}\frac{\vec{g}_\vec{k}\vec{g}_\vec{k}^\dagger}{z-\omega_\vec{k}}\,.
\end{equation}
Note that this definition of $\bm{\Sigma}(z)$ is consistent with the definition of the self-energy matrix in Sec.~\ref{sec:multipleimpurities} of the main text. Thus, the eigenenergies $E$ are solutions to
\begin{equation}
\label{eq:eigenenergy_equation}
    \mathrm{det}\, \bm{G}^{-1}(E)=0\,,
\end{equation}
where $\bm{G}(z)\equiv[z-\Delta-\bm{\Sigma}(z)]^{-1}$, and the amplitudes on the impurities $\vec{a}_\alpha$, which fully determine the associated eigenstate through Eq.~\eqref{eq:impurity_amplitudes}, can be
chosen as unit vectors in the kernel of $\bm{G}^{-1}(E)$. 
It is straightforward to show that the norm of the state is given by
\begin{equation}
\label{eq:eigenstate_norm}
\langle\phi_\alpha(E)|\phi_\alpha(E)\rangle=\vec{a}_\alpha^\dagger\left[1-\bm{\Sigma}'(E)\right]\vec{a}_\alpha\,,
\end{equation}
where $\bm{\Sigma}'(z) \equiv \partial_z\bm{\Sigma}(z)$. Similarly, the degenerate eigenstates obey the orthogonality condition
\begin{equation}
\label{eq:eigenstate_orthogonality}
    \langle\phi_\alpha(E)|\phi_\beta(E)\rangle=\vec{a}_\alpha^\dagger\left[1-\bm{\Sigma}'(E)\right]\vec{a}_\beta=0
\end{equation}
for $\alpha\neq\beta$.

Note that in the thermodynamic limit, the sum in Eq.~\eqref{eq:self-energy_matrix} diverges for an eigenenergy $E$ within the band of the bath, since the bath energies $\omega_\vec{k}$ form a continuum and so for some $\vec{k}$ in the sum, $\omega_\vec{k}\to E$, causing the denominator of the corresponding term to vanish. 
The norm of the eigenstates with energies in the continuum may diverge or not in the thermodynamic limit, depending on the specific values of $\vec{a}_{\alpha}$. The first kind correspond to scattering eigenstates, while the second correspond to bound states in the continuum (BIC), as discussed in the main text. In the following discussion, we disregard these states and focus only on the unbounded eigenstates.

We now consider the correlations of degenerate eigenstates in the initial state $\ket{\psi_0}=c^\dagger\ket{\rm FS}$. We find that
\begin{equation}
\label{eq:eigenstate_correlations}
\langle\phi^\dagger_\alpha(E)\phi_\beta(E)\rangle_0=\frac{\vec{a}_\alpha^\dagger\left[1-\bm{\Sigma}'_<(E)\right]\vec{a}_\beta}{\sqrt{\langle\phi_\alpha|\phi_\alpha\rangle\langle\phi_\beta|\phi_\beta\rangle}}
\end{equation}
where we have dropped the argument $E$ in the denominator, and where $\bm{\Sigma}_<(z)$ is defined analogously to Eq.~\eqref{eq:self-energy_matrix}, with the sum restricted to $\vec{k}\in\mathrm{BZ}$ such that $\omega_\vec{k} < E_F$. 
We want to understand the behaviour of this correlator in the thermodynamic limit. As noted above, the denominator in this expression will diverge. If $E>E_F$, then the divergent term which dominates $\bm{\Sigma}(E)$ is not included in $\bm{\Sigma}_<(E)$, and therefore the numerator in Eq.~\eqref{eq:eigenstate_correlations} remains finite and, accordingly, the correlator vanishes. On the other hand, if $E<E_F$, the numerator is dominated by the divergent term and hence $\vec{a}_\alpha^\dagger\left[1-\bm{\Sigma}'_<(E)\right]\vec{a}_\beta\sim\vec{a}_\alpha^\dagger\left[1-\bm{\Sigma}'(E)\right]\vec{a}_\beta$. By Eq.~\eqref{eq:eigenstate_orthogonality}, this implies that the correlator vanishes for $\alpha\neq\beta$. Conversely, for $\alpha=\beta$, the numerator and denominator diverge in the same way, so that finally
\begin{equation}
    \langle\phi^\dagger_\alpha(E)\phi_\beta(E)\rangle_0=\delta_{\alpha\beta}\Theta(E_F-E)\,,
\end{equation}
for unbounded eigenstates in the thermodynamic limit, which is what we set out to show.

\section{Population trapping in BICs}
\label{app:BICs}

\subsection{General formalism}
Before discussing in general terms the contribution of BICs to the LTA impurity occupation, we first review the formalism for BICs developed in Refs.~\cite{leonforte_vacancy-like_2021,leonforte_dressed_2021}. The bath Hamiltonian in Eq.~\eqref{eq:Hamiltonian} can be written as $H_B=H_B^\bullet+H_B^\circ$, where $H_B^\bullet$ includes only the sites to which the impurities couple while $H_B^\circ$ describes a bath with vacancies at those sites. A general BIC has the form $\ket{\phi}=\sum_{n}\psi_n\ket{e_n}+\ket{\Psi}$, where the bath component $\ket{\Psi}=\sum_j\psi_j\ket{j}$ obeys $H_B^\circ\ket{\Psi}=\Delta\ket{\Psi}$, i.e.\ it is an eigenstate of the bath with vacancies. Note that this implies, of course, that $\psi_{j_n}=0$. The emitter amplitudes are then given by 
\begin{equation}
\label{eq:BIC_impurity_amplitude}
    \psi_n=\frac{J}{g}\sum_{\langle j_n,j\rangle}\psi_j\,.
\end{equation}
Importantly, the BICs obtained in this way are not, in general, orthonormalised. Specifically, given a set of BICs $\{\ket{\phi_n}\}_n$, we can define a (non-unitary) transformation $\ket{\tilde{\phi}_\mu}=\sum_nM_{n\mu}\ket{\phi_n}$, such that Eq.~\eqref{eq:LTAvsGS_general} becomes, neglecting the contributions from the ground state and any other bound states,
\begin{multline}
\label{eq:formal_BIC_LTA}
\overline{\mean{O}}
    =\mathrm{tr}\Big[\Big(\bm{M}^\dagger \bm{O}\bm{M}\Big)\Big(\bm{M}^\dagger \bm{X}\bm{M}\Big)\Big] \\
    +\sum_{\omega_\vec{k}<E_F}\mathrm{tr}\Big[\Big(\bm{M}^\dagger \bm{O}\bm{M}\Big)\Big(\bm{M}^\dag\bm{Y}(\vec{k})\bm{M}\Big)\Big]\,.
\end{multline}
Here, we have also used the fact that all BICs are degenerate at energy $\Delta$ and defined matrices $\bm{O}$, $\bm{X}$ and $\bm{Y}(\vec{k})$ with elements
\begin{subequations}
\label{eq:BIC_matrices}
\begin{align}
&O_{mn}=\langle\phi_m|O|\phi_n\rangle \\
&X_{mn}\equiv\sum_{l=1}^N\langle\phi_m|e_l\rangle\langle e_l|\phi_n\rangle \\
&Y_{mn}(\vec{k})\equiv\langle\phi_m|\vec{k}\rangle\langle \vec{k}|\phi_n\rangle
\end{align}
\end{subequations}
By construction, $\langle\tilde{\phi}_\mu|\tilde{\phi}_\nu\rangle=\delta_{\mu\nu}$ implies that $\bm{M}^\dagger\bm{S}\bm{M}=I$, where $\bm{S}$ is the overlap matrix of the non-orthonormalised BICs with elements $S_{nm}=\langle\phi_n|\phi_m\rangle$. Since $\bm{S}$ is positive-semidefinite, a suitable choice of $\bm{M}$ is $\bm{M}=\bm{S}^{-1/2}=\bm{V}\bm{D}^{-1/2}\bm{V}^\dag$, where $\bm{V}$ diagonalises $\bm{S}$ and $\bm{D}=\mathrm{diag}(\lambda_1,\ldots,\lambda_{N-1})$ contains its eigenvalues.

\subsection{Dicke regime (1D)}

\subsubsection{BIC population trapping}
As alluded to in the main text, if a pair of impurities is separated at a distance $d=(\pi/k_\Delta)\mathbb{Z}$, the lattice described by $H_B^\circ$ consists of a finite chain of length $d-1$, as well as two semi-infinite chains with unbounded eigenstates (in the thermodynamic limit). The eigenmodes of the finite chain are  standing waves. The particular standing wave eigenmode at energy $\Delta$ thus forms the localised bath component of a BIC, and by Eq.~\eqref{eq:BIC_impurity_amplitude}, this BIC only has amplitudes on the two impurities which ``cut out'' the finite chain from the lattice. In particular, this implies that in the Dicke regime there emerge $N-1$ such BICs, whose explicit form can be calculated, after normalisation, as~\cite{busch_tight-binding_1987}
\begin{equation}
\label{eq:Dicke_BIC}
    \ket{\phi_n}=\sqrt{R}\left\{\ket{e_n^\pm}+\sqrt{\frac{\Gamma_0}{v_\Delta}}\sum_{j=1}^{d-1}\sin(k_\Delta j)\ket{j_n+j}\right\}\,,
\end{equation}
where $n=1,\ldots,N-1$, $\ket{e_n^\pm}=(\ket{e_n}\pm\ket{e_{n+1}})/\sqrt{2}$ with $\pm$ referring to the case where $\cos(k_\Delta d)=\mp1$, and $R$ defined in Eq.~\eqref{eq:Dicke_residue}.

We now focus on the case of the LTA total impurity occupation, for which $\bm{O}=\bm{X}$ in Eq.~\eqref{eq:formal_BIC_LTA}. Using the wavefunction~\eqref{eq:Dicke_BIC}, we can compute
\begin{subequations}
\label{eq:Dicke_matrices}
\begin{align}
    X_{mn} & = R \left[\delta_{m, n} \mp \frac{1}{2}\left(\delta_{m,n+1} + \delta_{m, n-1}\right)\right] \\
    Y_{mn}(k) & = \frac{g^2R}{L} \left[\frac{1\mp\cos(kd)}{(\Delta-\omega_k)^2}\right]\,e^{-ik(m-n)d} \\
    S_{mn} & = \delta_{m, n} \mp \frac{R}{2}\left(\delta_{m, n+1} + \delta_{m, n-1}\right)
\end{align}
\end{subequations}
where $\mp$ refers to $\cos(k_\Delta d)=\pm 1$. Since $\bm{S}$ is tridiagonal and Topelitz, its eigensystem can be obtained analytically and reads~\cite{noschese_tridiagonal_2013}
\begin{subequations}
\begin{align}
    &\lambda_n=1+R\cos\left(\frac{n\pi}{N}\right) \,, \\
    &V_{nm}=\sqrt{\frac{2}{N}}\sin\left(\frac{nm\pi}{N}\right)
\end{align}
\end{subequations}
with $n,m=1,\ldots,N-1$. Notably, $\bm{V}=\bm{V}^T=\bm{V}^\dag$ and this transformation will diagonalise any tridiagonal Topelitz matrix with equal values on the off-diagonals. In particular, this includes $\bm{X}$, which has eigenvalues
\begin{equation}
    \chi_n=R\left[1+\cos\left(\frac{n\pi}{N}\right)\right] \,,
\end{equation}
so that we can write
\begin{equation}
\label{eq:formal_BIC_eigs}
    \mathrm{tr}\Big[\Big(\bm{M}^\dagger \bm{X}\bm{M}\Big)^2\Big]=\sum_{m=1}^{N-1}\left(\frac{\chi_n}{\lambda_n}\right)^2
\end{equation}
which gives us Eq.~\eqref{eq:Dicke_LTA}. Similarly,
\begin{equation}
    \mathrm{tr}\Big[\Big(\bm{M}^\dagger \bm{X}\bm{M}\Big)\Big(\bm{M}^\dag\bm{Y}(k)\bm{M}\Big)\Big]=\sum_{m=1}^{N-1}\left(\frac{\chi_n}{\lambda_n^2}\right)(\bm{V}\bm{Y}(k)\bm{V})_{nn}
\end{equation}
and, after some algebra, we find that
\begin{multline}
\label{eq:formal_BIC_complicated}
    (\bm{V}\bm{Y}(k)\bm{V})_{nn}=\frac{g^2R}{L}\left[\frac{1\mp\cos(kd)}{(\Delta-\omega_k)^2}\right]\\ \times\frac{\left[1-(-1)^n\cos(Nkd)\right]\sin^2\left(\dfrac{n\pi}{N}\right)}{N\left[\cos\left(\dfrac{n\pi}{N}\right)-\cos(kd)\right]^2}\,.
\end{multline}

\subsubsection{Two-impurity resolvent formalism}
To draw a connection to previous work, we can confirm that our explicit analysis of population trapping in terms of BICs is consistent with the resolvent approach used to describe this effect in the single-excitation sector~\cite{gonzalez-tudela_markovian_2017,gonzalez-tudela_quantum_2017}. For two impurities, we consider (anti)symmetrised impurity operators $c_\pm=(c_1\pm c_2)/\sqrt{2}$, since for a time-reversal symmetric bath ($\omega_k=\omega_{-k}$) these couple to two sets of orthogonal bath modes with the same dispersion relation $\omega_k$ as the original bath. Specifically, for emitters at a distance $d$ from each other, $H = H_+ + H_-$ with
\begin{multline}
    H_\pm
    =\Delta c_\pm^\dagger c_\pm+\sum_{k>0}\omega_kb_{k,\pm}^\dagger b_{k,\pm} \\
    +\sum_{k>0}\frac{g_{k,\pm}}{\sqrt{L}}\left(c_\pm^\dagger b_{k,\pm}+h.c.\right)
\end{multline}
with coupling $g_{k,\pm}=g\sqrt{2(1\pm\cos kd)}$ and bath modes
\begin{equation}
    b_{k,\pm}=\frac{\left(e^{ikr_1}\pm e^{ikr_2}\right)b_k+\left(e^{-ikr_1}\pm e^{-ikr_2}\right)b_{-k}}{2\sqrt{1\pm\cos kd}}
\end{equation}
which satsify $\{b_{k,\alpha}^\dagger,b_{q,\beta}\}=\delta_{\alpha\beta}\delta_{kq}$ for $k\in[0,\pi)$. Since $[H_+,H_-]=0$, the two-impurity problem can thus be mapped onto two single-impurity problems.

To extend the resolvent formalism to each of these problems, we define the states $\ket{e_\pm}\equiv c_\pm^\dagger\ket{\rm vac}$ and $\ket{k_\pm}\equiv b_{k,\pm}^\dagger\ket{\rm vac}$ and the transition amplitudes $\mathcal{A}_{\pm}(t)\equiv\bra{e_\pm}e^{-iHt}\ket{e_\pm}$ and $\mathcal{A}_{k,\pm}(t)\equiv\bra{k_\pm}e^{-iHt}\ket{e_\pm}$, which can be calculated from an integral of the form~\eqref{eq:resolvent_integral} from propagators $G_\pm(z)$ and $G_{k,\pm}(z)$ identical to $G_e(z)$ and $G_k(z)$ in Eqs.~\eqref{eq:propagators} except for the substitution $\Sigma_e(z)\to\Sigma_\pm(z)\equiv\Sigma_e(z) \pm \Sigma(z, d)$ and $g\to g_{k,\pm}$. By a derivation analogous to the one for the single-impurity model, the collective decay of the fully inverted state is described by
\begin{equation}
    \mean{N_e}_t=\sum_{\alpha=\pm}\left(\abs{\mathcal{A}_\alpha(t)}^2+\sum_{k<k_F}\abs{\mathcal{A}_{k,\alpha}(t)}^2\right)\,.
\end{equation}
The long-term limit of this expression is determined by the real poles of $G_\pm(z)$ and $G_{k,\pm}(z)$. This again includes BS outside the energy band, given by the real solutions to the pole equations $z-\Delta-\Sigma_\pm(z)=0$, however we neglect their contribution here. This leaves only the pole $\omega_k$ of $G_{k,\pm}(z)$, as well as a new pole at $z=\Delta$, which we can show emerges for $G_\mp(z)$ when $\cos(k_\Delta d)=\pm 1$. The associated residue is given by $R$ as defined in Eq.~\eqref{eq:Dicke_residue}~\cite{gonzalez-tudela_markovian_2017}, such that the LTA occupation is finally given by
\begin{equation}
\begin{split}
    &\overline{\mean{N_e}}=g^2\int_{\omega_k<E_F}\frac{dk}{2\pi}\abs{G_e(\omega_k^+)}^2 \\
    &\qquad+R^2\left(1+g^2\int_{\omega_k<E_F}\frac{dk}{2\pi}\frac{1\mp\cos(kd)}{(\Delta-\omega_k)^2}\right)\,.
\end{split}
\end{equation}
Given Eq.~\eqref{eq:formal_BIC_complicated}, it is straightforward to see that this agrees with Eq.~\eqref{eq:formal_BIC_LTA} in the case where $N=2$.

\subsection{Van Hove singularity (2D)}
When searching for BICs in the 2D bath with impurities detuned to the Van Hove singularity, we are faced with the challenge that the entries of the self-energy matrix $\bm{\Sigma}(\Delta)$ will, in general, diverge at $\Delta=4J$. We should therefore understand Eq.~\eqref{eq:BIC_condition} as the statement that $\bm{\Sigma}(\Delta)$ has at least one eigenvalue which vanishes in the limit $\Delta\to 4J$. However, rather than trying to establish which constraints this condition imposes on the impurity positions in general, we note that there is a more intuitive approach for constructing BICs at the Van Hove singularity.

\begin{figure}
    \centering
    \includegraphics[width=\linewidth]{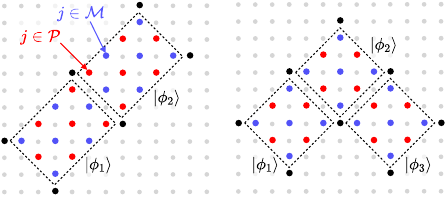}
    \caption{BIC wavefunctions for the extended emitter configurations from Fig.~\ref{fig:BIC_decay}b, uniquely defined up to a global sign. Black dots indicate sites to which the impurities are coupled (i.e.\ vacancy sites) and we highlight a positive (negative) non-zero amplitude $\psi_j$ on a bath site $j$ in red (blue).}
    \label{fig:BIC_wavefunctions}
\end{figure}

For such a BIC, it is easy to see that the eigenvalue equation $H_B^\circ\ket{\Psi}=\Delta\ket{\Psi}$ amounts to the condition
\begin{equation}
\label{eq:VHS_BIC_condition}
    \sum_{\langle j,j'\rangle}\psi_{j'}=0~~\forall~~j\neq j_n\,.
\end{equation}
We can then use Eq.~\eqref{eq:VHS_BIC_condition} to iteratively construct a simple, real wavefunction for the BIC supported by this configuration: starting at some site $j$ adjacent to a vacancy site, we assign $\psi_j=+1$ or $\psi_j=-1$ arbitrarily on this site. To ensure that Eq.~\eqref{eq:VHS_BIC_condition} holds for the nearest neighbours of this site, we must then assign amplitudes $\psi_{j'}=\pm1$ to its next-nearest neighbours $j'$. We then proceed to enforce Eq.~\eqref{eq:VHS_BIC_condition} to those sites in turn, until we arrive ultimately at a checkerboard pattern, where every other site in the rectangle has amplitude $\psi_j=+1$ (which we label $j\in P$) or $\psi_j=-1$ ($j\in M$). After normalisation, we are left with a BIC wavefunction
\begin{equation}
\label{eq:VHS_BIC}
    \ket{\phi}=\sqrt{R_V}\left\{\ket{\bm{\sigma}}+\frac{g}{2J}\left(\sum_{j\in P}\ket{j}-\sum_{j\in M}\ket{j}\right)\right\}\,,
\end{equation}
where we have defined
\begin{equation}
    R_V=\left[1+\left(\frac{g}{2J}\right)^2\left(\abs{P}+\abs{M}\right)\right]^{-1}\,,
\end{equation}
with $\abs{P}$ and $\abs{M}$ denoting the number of sites with positive and negative amplitude, respectively, and where
\begin{equation}
    \ket{\bm{\sigma}}=\frac{1}{2}\sum_{n=1}^4\sigma_n\ket{e_n}\,,
\end{equation}
with $\bm{\sigma}\in\{-1,+1\}^{\otimes 4}$ capturing the relative sign of each emitter amplitude as determined from the bath wavefunction by Eq.~\eqref{eq:BIC_impurity_amplitude}.

With reference to Refs.~\cite{gonzalez-tudela_markovian_2017,gonzalez-tudela_quantum_2017}, we note that the above analysis applies for instance to the diamond configuration of four impurities at sites $\vec{r}_{1}=(0,-m)$, $\vec{r}_2=(-m,0)$, $\vec{r}_3=(0,m)$, and $\vec{r}_4=(m,0)$ (shown for $m=3$ in Fig.~\ref{fig:BIC_wavefunctions}a). In this case, $\abs{P}+\abs{M}=m^2$ and therefore $R_V=\left[1+(gm)^2/(2J)^2\right]^{-1}$. Noting also that $\bm{\sigma}=(1,\pm1,1,\pm1)$, where $+$ ($-$) refers to the case of odd (even) $m$. It is then easy to show that the subradiant decay in this configuration reported in Refs.~\cite{gonzalez-tudela_markovian_2017,gonzalez-tudela_quantum_2017} is consistent with population trapping in a BIC of the form~\eqref{eq:VHS_BIC}.

More generally, our analysis can also be applied to extended configurations giving rise to multiple BICs of the form~\eqref{eq:VHS_BIC}. We focus in particular on the two examples shown in Fig.~\ref{fig:BIC_decay}, with $N=6$ and $N=8$ impurities, respectively, whose wavefunctions can be constructed shown in Fig.~\ref{fig:BIC_wavefunctions}. It is straightforward to show that these configurations support multiple BICs $\ket{\phi_n}$ with $n=1,2$ and $n=1,2,3$, respectively. With access to the full BIC wavefunction, we are able to construct the matrices $\bm{S}$, $\bm{O}$, $\bm{X}$ and $\bm{Y}(\vec{k})$ explicitly, allowing us to calculate LTA expectation values from Eq.~\eqref{eq:formal_BIC_LTA}.

\end{document}